\documentclass[letterpaper,11pt]{article}
\usepackage{caption}
\pdfoutput=1 
\usepackage{xcolor}
\usepackage{enumerate}
\usepackage{framed}
\usepackage{mdframed}
\usepackage{amsmath}
\usepackage{amsfonts}
\usepackage{mathrsfs}
\usepackage{euscript}
\usepackage{amssymb}
\usepackage[normalem]{ulem}
\usepackage{graphicx, rotating}
\usepackage{epsfig}
\usepackage{latexsym}
\usepackage{graphicx}
\usepackage{color}
\usepackage{amsmath,bm,amssymb}
\usepackage{cite}
\usepackage{slashed}
\usepackage{hyperref}
\usepackage{epstopdf}
\usepackage[title]{appendix}
\usepackage[font=footnotesize,labelfont=]{caption}
\AppendGraphicsExtensions{.tif}
\hypersetup{colorlinks=true, citecolor=bluscuro, linkcolor=black, urlcolor=bluscuro}
\definecolor{rossos}{cmyk}{0,1,1,0.55}
\definecolor{bluscuro}{rgb}{0.15, 0.2, .85}
\definecolor{bluchiaro}{cmyk}{1,.3,0.,0.1}

\newcommand{\be}{\begin{equation}}
\newcommand{\ee}{\end{equation}}
\newcommand{\bea}{\begin{eqnarray}}
\newcommand{\eea}{\end{eqnarray}}
\newcommand{\beas}{\begin{eqnarray*}}
\newcommand{\eeas}{\end{eqnarray*}}

\newcommand{\pvec}[1]{\vec{#1}\mkern2mu\vphantom{#1}}
\def\BH{\text{\tiny BH}}
\def\DM{\text{\tiny DM}}
\def\DF{\text{\tiny DF}}

\newcommand{\lp}{\left (}
\newcommand{\rp}{\right )}
\def\d{{\rm d}}

\begin{document}
\def\thefootnote{\fnsymbol{footnote}}

\begin{center}
\LARGE{\textbf{Dynamical friction in dark matter superfluids:\\
The evolution of black hole binaries}} \\[0.5cm]

\large{{Lasha Berezhiani,$^{1,2}$\footnote{lashaber@mpp.mpg.de} Giordano Cintia,$^{1,2}$\footnote{cintia@mpp.mpg.de} Valerio De Luca,$^{3}$\footnote{vdeluca@sas.upenn.edu}  and Justin Khoury$^{\, 3}$\footnote{jkhoury@sas.upenn.edu}}
}\\[0.5cm]

\small{
\textit{
$^1$Max-Planck-Institut f\"ur Physik, F\"ohringer Ring 6, 80805 M\"unchen, Germany\\
 \vskip 5pt
~~$^2$ Arnold Sommerfeld Center, Ludwig-Maximilians-Universit\"at, \\Theresienstra{\ss}e 37, 80333 M\"unchen, Germany\\
 \vskip 5pt
~$^3$Center for Particle Cosmology, Department of Physics and Astronomy, \\ University of Pennsylvania, 209 South 33rd St, Philadelphia, PA 19104, USA}
 }

 \vspace{.2cm}

\end{center}

\vspace{.6cm}

\hrule \vspace{0.2cm}
\centerline{\small{\bf Abstract}}
\vspace{0.1cm}
{\small\noindent 
The theory of superfluid dark matter is characterized by self-interacting sub-eV particles that thermalize and condense to form a superfluid core in galaxies.
Massive black holes at the center of galaxies, however, modify the dark matter distribution and result in a density enhancement in their vicinity known as dark matter spikes. The presence of these spikes affects the evolution of binary systems by modifying their gravitational wave emission and inducing dynamical friction effects on the orbiting bodies. In this work, we assess the role of dynamical friction for bodies moving through a superfluid core enhanced by a central massive black hole. As a first step, we compute the dynamical friction force experienced by bodies moving in a circular orbit. Then, we estimate the gravitational wave dephasing of the binary, showing that the effect of the superfluid drag force is beyond the reach of space-based experiments like LISA, contrarily to collisionless dark matter, therefore providing an opportunity to distinguish these dark matter models. 
}

\vspace{0.3cm}
\noindent
\hrule
\def\thefootnote{\arabic{footnote}}
\setcounter{footnote}{0}

\section{Introduction} 

The standard~$\Lambda$ Cold Dark Matter ($\Lambda$CDM) model, wherein dark matter (DM) consists of non-relativistic, collisionless particles, offers an exquisite fit to various observations across different range of scales, from galaxies to large-scale structures of the Universe. Whether this empirical success persists down to galactic scales, where baryons play an important role in shaping the DM distribution ({\it e.g.},~\cite{Oman:2015xda}), remains hotly debated~\cite{Bullock:2017xww}. Alternatively, these potential small-scale shortcomings may indicate that an extension of the model is warranted, for example 
endowing DM with self-interactions that are strong enough to modify the galactic core~\cite{Spergel:1999mh,Kaplinghat:2015aga}. Another well-motivated extension is to consider ultralight DM particles, called fuzzy dark matter~\cite{Hu:2000ke,Hui:2016ltb}, which have a de Broglie wavelength of order kiloparsec and behave as a Bose-Einstein condensate on astrophysical scales~\cite{Ferreira:2020fam}.

Yet another possibility is that DM consists of self-interacting, sub-eV bosonic particles that can achieve superfluidity. Over a certain mass range and scattering length, these particles can generate a condensate in sufficiently dense environments and low enough temperature~\cite{Goodman:2000tg}. 
In some realizations of the model, long-range interactions between baryons can be mediated by phonons, thereby affecting galactic dynamics of ordinary matter~\cite{Berezhiani:2015pia,Berezhiani:2015bqa,Berezhiani:2017tth,Sharma:2018ydn,Berezhiani:2018oxf}. 
The simplest superfluid theory, based on quartic self-interactions, was initially ruled out in Ref.~\cite{Slepian:2011ev} from Bullet Cluster constraints and observations of galactic rotation curves~\cite{Markevitch:2003at,Clowe:2003tk}, while recent weak-lensing observations constrain the model when a Modified Newtonian Dynamics (MOND)-like behaviour on galactic scales is approached thanks to a direct coupling to baryons~\cite{Mistele_2023}. However, it has been recently revitalized in Refs.~\cite{Berezhiani:2021rjs,Sharma:2022jio,Berezhiani:2022buv}, by relaxing the assumptions of global thermal equilibrium and spherical symmetry.

It has been long established that the DM distribution affects the galaxy formation history~\cite{Blumenthal:1984bp,2008gady.book.....B}. Density profiles obtained from numerical fits or~$N$-body cosmological simulations, such as the Navarro-Frenk-White (NFW) profile~\cite{Navarro:1996gj}, suggest that the mass distribution is peaked near the galactic center and dies off at larger distances. The predicted distribution is, however, affected in the presence of a central black hole (BH), depending both on the properties of the DM particles and the formation history of the central object. In an early study, Gondolo and Silk developed a Newtonian scheme to study how collisionless DM redistributes around the BH starting from an initial power-law distribution~\cite{Gondolo:1999ef}. They showed the formation of a spike, with a density peak near the radius of marginally bound circular orbits, below which the DM particles are accreted into the BH. The analysis has been later improved in Ref.~\cite{Sadeghian:2013laa}, where a fully general relativistic calculation showed that the spikes can get significantly higher and closer to the central BH.
On the other hand, the gravitational scattering of stars in the inner region can heat up the DM fluid and give rise to a softened profile~\cite{Merritt:2003qk, Gnedin:2003rj, Merritt:2006mt,Shapiro:2022prq} or even to disruption~\cite{Wanders:2014xia}, while DM annihilations may induce a smoother cusp~\cite{Vasiliev:2007vh,Shapiro:2016ypb}. Similarly, the presence of DM self-interactions~\cite{Fornasa:2007nr,Shapiro:2014oha,Feng:2021qkj, Chavanis:2019bnu} may wash out any initial and/or adiabatically altered particle distribution and give rise to smoother spikes~\cite{Shapiro:2014oha}.

Recently, two of us calculated the density profile of superfluid DM around supermassive BHs assuming different superfluid equations of state, for example when it is predominantly characterized by two-body and three-body interactions~\cite{DeLuca:2023laa}. It was found that, depending on the distance from the central object,
the DM density is characterized by different functional behaviors: at large distances the DM appears as a superfluid core, which disappears into a cusp as we approach smaller distances, with an overall enhancement by a factor of order~$10^2-10^3$ relative to the asymptotic core density. The characteristics of this profile 
distinguish it from the standard predictions for collisionless DM~\cite{Bertschinger:1985pd}.

The formation of these superfluid spikes may have important consequences for the formation of binaries near the galactic cores, where a compact object can become gravitationally bound to the central BH. These binaries have a mass ratio that spans values of order~$10^{-4}$ to~$10^{-3}$ for intermediate mass-ratio inspirals~\cite{Amaro-Seoane:2007osp}, and~$10^{-6}$ to~$10^{-5}$ for extreme mass-ratio inspirals~\cite{Babak:2017tow}. They can be observed with upcoming third-generation gravitational wave (GW) experiments like the Einstein Telescope~\cite{Punturo:2010zz,Maggiore:2019uih,Branchesi:2023mws} and Cosmic Explorer~\cite{Reitze:2019iox}, or space-based interferometers like LISA~\cite{LISA:2017pwj}, respectively.
The impact of a DM spike on these binaries is twofold: it modifies the gravitational well felt by the secondary compact object, and it induces dynamical friction drag forces that slow down the binary's trajectory~\cite{Barausse:2014tra}.

Dynamical friction is a phenomenon wherein an object (perturber) moving through a discrete or continuous medium generates a density fluctuation (wake) in that medium.
The gravitational attraction from this wake has the effect of  slowing down the motion of the object.
The pioneering study of Chandrasekhar~\cite{Chandrasekhar:1943ys} considered a perturber linearly moving in a collisionless medium, while Refs.~\cite{1971ApJ...165....1R, 1980ApJ...240...20R, Ostriker:1998fa} focused on linear motion in a gaseous medium.
The case of circular binaries was also investigated in Refs.~\cite{Kim:2008ab, 2009ApJ...703.1278K, Lee:2011px, Lee:2013wtc,2022ApJ...928...64D,Buehler:2022tmr}.
More diverse environments, such as fuzzy dark matter, self-interacting or superrandiance-driven scalar clouds, and Bose-Einstein condensates, have also been explored recently~\cite{Hui:2016ltb,Lancaster:2019mde, Annulli:2020lyc, Hartman:2020fbg, Wang:2021udl,Boudon:2022dxi,Boudon:2023qbu,Baumann:2021fkf,Tomaselli:2023ysb}.

For the case of superfluid DM, Ref.~\cite{Berezhiani:2019pzd} computed the dynamical friction force experienced by a perturber moving through a uniform core either at constant velocity or accelerated due to an external field. For the steady motion, it highlighted the relevance of higher-gradient corrections in the phonon dispersion relation, which are responsible for the so-called quantum pressure in the hydrodynamical description. Our goal in this work is to understand the impact of a density spike on the evolution of a binary system orbiting within a superfluid environment. To this end, we compute the dynamical friction force experienced by a small perturber moving in circular orbits around a massive black hole surrounded by a DM spike, and then discuss the relevance of the drag force relative to the emission of GWs.

The paper is organized as follows. In Sec.~\ref{Sec2} we review the basic notion of the superfluid DM model and the corresponding spikes around BHs. In Sec.~\ref{Sec3} we calculate the dynamical friction force for circular binaries. In Sec.~\ref{Sec4} we put these results together to discuss the evolution of black hole binaries. We conclude and summarize our results in Sec.~\ref{Sec5}. Two Appendices are devoted to technical details. In the following, we use natural units~$\hbar = c = 1$.

\section{Excursus on superfluid dark matter} 
\label{Sec2}
In this Section, we summarize the basic description of the theory of superfluid DM~\cite{Berezhiani:2015pia,Berezhiani:2015bqa,Berezhiani:2017tth,Sharma:2018ydn,Berezhiani:2021rjs,Sharma:2022jio, Berezhiani:2022buv}. In this model, DM particles are sufficiently light and strongly self-interacting to thermalize and condense at the center of galaxies, creating a superfluid core.

\subsection{Theory of dark matter superfluidity}

The simplest realization of the superfluid DM model consists of a non-relativistic massive complex scalar field~$\psi$, with mass~$m_\DM$ and quartic self-interactions, described by the Hamiltonian:
\begin{equation}
\mathcal{H}_2 = \int \text{d}^3x \left[-\frac{1}{2m_\DM} \psi^{\dagger}(x)\vec{\nabla}^2 \psi(x) + \frac{1}{2}g_2 \,\psi^{\dagger\, 2}(x) \psi^2(x)  \right]\,.
\label{H2}
\end{equation}
We assume that this bosonic sector is minimally-coupled to gravity. The strength of the contact interactions is parameterized by the coupling~$g_2$, which is responsible for  Bose-Einstein condensation and for generating a stable superfluid state, if it is positive,~$g_2>0$. In other words, repulsive self-interactions are essential for the system to exhibit superfluidity. The theory has a global U(1) symmetry, with the corresponding Noether charge given by the particle number
density~$n$.

From an effective field theory perspective, the superfluid condensate is described by a classical field configuration with finite number density, which spontaneously breaks the U(1) global symmetry. At zero temperature, the superfluid equation of state is given by
\begin{equation}
\label{EOS2}
P_2=\frac{g_2 \rho^2}{2m^2_{\DM}}\,, 
\end{equation}
in terms of the mass density~$\rho = m_\DM n$.
 The corresponding spectrum of gapless perturbations ({\it i.e.}, phonons) around the homogeneous solution is given by~\cite{Chavanis:2011zi,Berezhiani:2019pzd}
\be
\label{dispersion2body}
\omega_k^2 = - 4 \pi G \rho + c_s^2 k^2 + \frac{k^4}{4 m_\DM^2}\,,
\ee
where
\be
c_s^2 = 
\frac{g_2\rho}{m^2_\DM}
\ee
is the adiabatic sound speed of the superfluid, corresponding to the first derivative of the pressure with respect to density.

The non-relativistic approximation inherent in Eq.~\eqref{H2} is valid as long as~$c_s \ll 1$. The first term in Eq.~\eqref{dispersion2body} describes the tachyonic contribution associated with the system's gravitational instability; the second term is the energy cost to excite a sound wave of momentum~$k$; and the third term represents the kinetic energy~$\frac{k^2}{2m_\DM}$ of a constituent of the system.
In the following, we ignore the thermal corrections to the equation of state and sound speed, proportional to the temperature of the system~\cite{Sharma:2018ydn}.

From the above dispersion relation, one can show that modes softer than the Jeans scale~$k_{\rm J}$ are unstable~\cite{Chavanis:2011zi,Berezhiani:2022buv}
\be
k^2_{\rm J} = 2 m_\DM^2 c_s^2 \lp - 1 + \sqrt{1 + \xi^{-1}} \rp\, \qquad \text{with} \qquad \xi=\frac{m_\DM^2 c_s^4}{4 \pi G \rho}.
\ee
The corresponding Jeans length scale,~$\lambda_{\rm J} = 2 \pi/ k_{\rm J}$, sets an upper bound on the size of a gravitationally stable superfluid core. 
One can identify two different regimes depending on the magnitude of the dimensionless parameter $\xi$,
which encodes the nature of the pressure sustaining the core.

\begin{itemize}

\item For negligible self-interactions,~$\xi \ll 1$, the ``quantum pressure" of the system prevents the gravitational collapse, giving rise to a Jeans scale of
\be
\lambda_{\rm J} \simeq \lp \frac{\pi^3}{G \rho m_\DM^2} \rp^{1/4}, \qquad \xi \ll 1\,.
\ee
This is the Jeans scale which is characteristic of fuzzy DM models, a regime we are not considering in this paper.

\item On the other hand, in the regime~$\xi \gg 1$ self-interactions are strong enough to be the main contribution in stabilizing against the collapse of the superfluid. In this case, the superfluid core is sustained by the \textit{interaction pressure}, with the corresponding Jeans scale
\be
\lambda_{\rm J} \simeq \sqrt{\frac{\pi c_s^2}{G \rho}}, \qquad \xi \gg 1\,.
\label{jsIP}
\ee
This is the regime of interest to describe a superfluid state, and we henceforth refer to it as \textit{interaction pressure case}. 

\end{itemize}

Let us mention that different theories can be considered as well.
For example, one can also envisage a theory beyond quartic self-interactions, where the superfluids are mostly described by three-body interactions~\cite{Berezhiani:2015bqa}. The corresponding interaction Hamiltonian would involve a hexic potential of the form
\begin{equation}
  \mathcal{H} \supset  \frac{1}{3}g_3\, \psi^{\dagger\, 3}(x) \psi^3(x)\,,
\end{equation}
in terms of the coupling constant~$g_3$.
The corresponding zero-temperature equation of state and sound speed read
\begin{equation}
\label{EOS3}
P_3
=  \frac{\rho^3}{12\Lambda^2m_\DM^6}\,; \qquad 
c_s^2 =
\frac{\rho^2}{4\Lambda^2m_\DM^6}\,,
\end{equation}
in terms of the cutoff scale~$\Lambda = 1/\sqrt{8 g_3 m_\DM^3}$.
In the non-relativistic limit, the phonon dispersion relation for the three-body case is equivalent to the one for quartic interactions, shown in Eq.~\eqref{dispersion2body}. The interested reader can find additional details in Appendix \ref{AppA}.

\subsection{Formation of the superfluid phase}

Self-interacting bosons can generate a superfluid state through Bose-Einstein condensation, which occurs if the conditions of {\it degeneracy} and {\it thermalization} are satisfied. However, it is important to clarify the role of these two conditions in forming the dark matter superfluid phase. 

Degeneracy occurs when the de Broglie volume of the system is highly occupied, and it can be satisfied also by systems which are out-of-equilibrium. Examples are models of fuzzy dark matter or cosmological axions, in which self-interactions are never sufficiently efficient to establish equilibrium. In general, one can define the coherence scale~$\lambda_{\rm c}$ of the system, describing the scale below which the gas of DM particles can be considered as an effective condensate. This is set by the minimum between the scale of gravitational stability and the de Broglie wavelength~\cite{Guth:2014hsa}
\begin{equation}
\lambda_{\rm c}\simeq \min\left(\lambda_{\rm J},\lambda_\text{\tiny dB}\right)\qquad \text{with}\qquad \lambda_\text{\tiny dB}=\frac{2\pi}{m_\DM v}\,. 
\label{ch3:coherencescale}
\end{equation} 
Here,~$v$ is the velocity dispersion of the system. Let us investigate what properties DM should have to determine a coherence length of order kpc.
Without thermalization (so without a mechanism that would push all particles in the zero momentum state), the typical velocity dispersion of DM particles in galaxies is~$v\simeq 10^{-3}$ and extremely light masses ($m_\DM \simeq 10^{-21}$~eV)  are required to have coherence over kpc scales.
In this case the de Broglie wavelength is comparable to the Jeans length, and a coherent, kpc-size, degenerate soliton is formed.

On the other hand, if one focuses on heavier DM particles, it is the de Broglie wavelength that sets the scale of coherence, with sub-kpc extent. For these masses, in order to obtain a superfluid state which is coherent over the Jeans length and extends to kpc scales,
both degeneracy and equilibrium are necessary (the latter condition responsible for decreasing the velocity dispersion of the system).
These requirements can be translated into bounds on the parameters of the model. 

The condition of {\it degeneracy} requires that the de Broglie wavelengths~$\lambda_\text{\tiny dB}$  of the particles overlap in galaxies. By requiring that~$\lambda_\text{\tiny dB}$ is larger than the average interparticle separation, one gets an upper bound on the DM mass~\cite{Berezhiani:2015pia,Berezhiani:2015bqa}
\be
m_\DM \lesssim \left(\frac{M_\text{\tiny DM}}{10^{12}M_\odot}\right)^{-1/4}\; {\rm eV}\,,
\ee
in terms of the DM halo mass~$M_\text{\tiny DM}$ at virialization.
This condition therefore implies that only light enough particles form a Bose-Einstein condensate, while heavy particles do not. 
This inequality can be milder if one considers halos that are only partially degenerate, but we will not be concerned with this possibility.

The second requirement, DM {\it thermalization} within galaxies, is achieved when the
bosonic particles reach their maximal entropy state via self-interactions. Central regions of the halo, characterized by higher densities with respect to its outskirts, are expected to catalyze thermalization, by enhancing interaction rates and facilitating equilibration. 
This defines a second important length scale, the thermal radius~$R_{\rm T}$.
This is the radius within which dark matter particles experienced at least one interaction over the galaxy lifetime~$t_g$. 
Therefore, the region of the halo which is in thermal equilibrium is defined by the condition
\begin{equation}
    \Gamma(R_{\rm T}) t_g=1\,,\qquad \text{with} \qquad \Gamma=\left(1+\mathcal{N}\right)\frac{\sigma}{m_\DM}\rho v\,.
    \label{RT}
\end{equation}
Here,~${\sigma} = m^2_\DM g_2^2/4\pi$ is the two-body scattering cross-section, while~$\rho$ is the density of the halo, which is assumed to be NFW-like. We also defined the Bose-enhancement factor,~$\mathcal{N}=n\lambda^3_\text{\tiny dB}$. If the gas is degenerate,~$\mathcal{N}\gg 1$, then interactions are enhanced by the highly occupied phase space.\footnote{If the gas is in equilibrium, the equipartition theorem for ideal gases allows to rewrite the condition~$\mathcal{N}\gg 1$ as
\begin{equation}
T \ll  T_{\rm c} = \frac{2\pi}{m_\DM^{5/3}}\left(\frac{\rho}{\zeta(3/2)}\right)^{2/3}\,,
\label{T_c ideal}
\end{equation}
where~$T_c$ is the critical temperature of the superfluid, and~$\zeta$ the Riemann zeta function.} The thermal radius enters in the non-linear equation \eqref{RT} through the density profile and the velocity.

To summarize, if sub-eV dark matter particles are considered, then the halo is degenerate and a macroscopically large number of particles occupy the ground state and give rise to a Bose-Einstein condensate in the region within~$R_{\rm T}$. Of course, if the thermal radius is larger than the virial radius itself, the whole halo will undergo the superfluid phase transition. We discuss these two scenarios in more detail in the next section.  

The self-interaction cross section is also subject to an upper bound from merging galaxy clusters, such as the Bullet Cluster~\cite{Markevitch:2003at, Clowe:2003tk}. In such merging events, 
the gas component is observed to be displaced with respect to the DM component. This observation requires that the DM particles must experience a negligible amount of scatterings while passing through the target cluster, giving rise to the bound
\be
\frac{\sigma}{m_\DM} \lesssim \left(1+\mathcal{N}\right)^{-1}\frac{{\rm cm}^2}{{\rm g}}\,,
\label{bullet}
\ee
under the assumption of 2-body scattering. Here~$\mathcal{N}$ is evaluated for typical cluster density distributions. Notice that this is not the ``standard" Bullet Cluster bound~$\frac{\sigma}{m_\DM} \lesssim \frac{{\rm cm}^2}{{\rm g}}$~\cite{Markevitch:2003at}. As discussed in \cite{Berezhiani:2021rjs,Berezhiani:2022buv}, the interaction rate is enhanced by the Bose-enhancement factor for degenerate particles, making the bound way more stringent. If we consider particles heavier than the eV scale, the bound relaxes to the well-known Bullet Cluster bound. It is important to emphasize that, as demonstrated in \cite{Berezhiani:2022buv}, this enhanced bound reverts to the standard non-degenerate counterpart also if the majority of dark matter in the halo is in the superfluid phase (i.e. when thermalization is also achieved). This is because the derivation of \eqref{bullet} relies on the assumption that particles are generally scattered into highly occupied energy levels. However, if most of the dark matter is in the condensed phase, this assumption no longer holds, and the use of~$\mathcal{N}$ becomes invalid.

\subsection{Superfluids in galaxies}

Once the conditions of degeneracy and thermalization are fulfilled, a superfluid phase of size~$R_{\rm T}$ is generated at the center of the dark matter halo. As long as the interaction pressure dominates over the degeneracy pressure, we have
\begin{equation}
    R_{\rm T}>\lambda_{\rm J}\,,\qquad (\xi\gg1)
\end{equation}
implying that the superfluid phase is, in general, susceptible to Jeans instability. The net result is that the core of size~$R_{\rm T}$ fragments into self-gravitating superfluid configurations.

The profile of these lumps can be computed by solving the hydrostatic equilibrium equation under the assumption of spherical symmetry. Using the equation of state in~Eq.~\eqref{EOS2}, the solution to the coupled Jeans-Poisson equations is 
\be
\rho(r) =  \rho_0 \frac{{\rm sin}\lp 2 \pi r/\lambda_{\rm J}\rp}{2 \pi r/\lambda_{\rm J}}\,,
\label{rho core2}
\ee
in terms of the core central density~$\rho_0$, and diameter~$\lambda_{\rm J}$. Notice that in the case of a two-body interacting superfluid, the latter coincides with the Jeans scale in the interaction pressure case \eqref{jsIP}. To have an idea of its dependence on the parameters of the model, we can rewrite it in terms of the scattering cross section as
\be
{\lambda_{\rm J}} 
\simeq 40 \, {\rm kpc} \lp \frac{m_\DM}{{\rm \mu eV}} \rp^{-5/4} \lp \frac{\sigma/m_\DM}{10^{-8} \, {\rm cm^2/g}} \rp^{1/4}\,.
\label{R2}
\ee
Let us stress that, in order to focus on the regime~$\lambda_{\rm J} \gtrsim 2$ kpc, we have to consider dark matter halos which are mostly in the superfluid phase. In this way, the enhanced bullet cluster bound \eqref{bullet} relaxes to its non-degenerate analog. If this were not the case, these values of the Jeans scale would be in conflict with the enhanced bound and  therefore be excluded.\footnote{For the three-body interacting case, with the equation of state given by Eq.~\eqref{EOS3}, the core profile takes the form~\cite{Berezhiani:2015bqa}
\be
\rho(r)\simeq \rho_0 \cos^{1/2}\left(\frac{\pi r}{2 R_3}\right)\,; \qquad 
R_3 = 2.75 \sqrt{\frac{\rho_0}{32\pi G \Lambda^2 m_\DM^6}}  \simeq 44 \, {\rm kpc} \lp \frac{m_\DM}{\rm eV} \rp^{-3} \lp \frac{\Lambda}{\rm meV} \rp^{-1} \lp \frac{\rho_0}{10^{-24} {\rm g/cm^3}} \rp^{1/2}\,.
\label{rho core3}
\ee
Notice that, in contrast to the two-body case, the soliton size~$R_3$ depends on the central density~$\rho_0$.
Following Ref.~\cite{Berezhiani:2015bqa}, throughout the paper we assume the fiducial values~$m_\DM = {\rm eV}$ and~$\Lambda = {\rm meV}$ for the model's parameters, which are possibly in tension with the Bullet cluster bound \cite{Berezhiani:2022buv}, but show indicatively how the picture changes for different equations of state.} 

\begin{figure}[t!]
	\centering
	\includegraphics[width=0.5\textwidth]{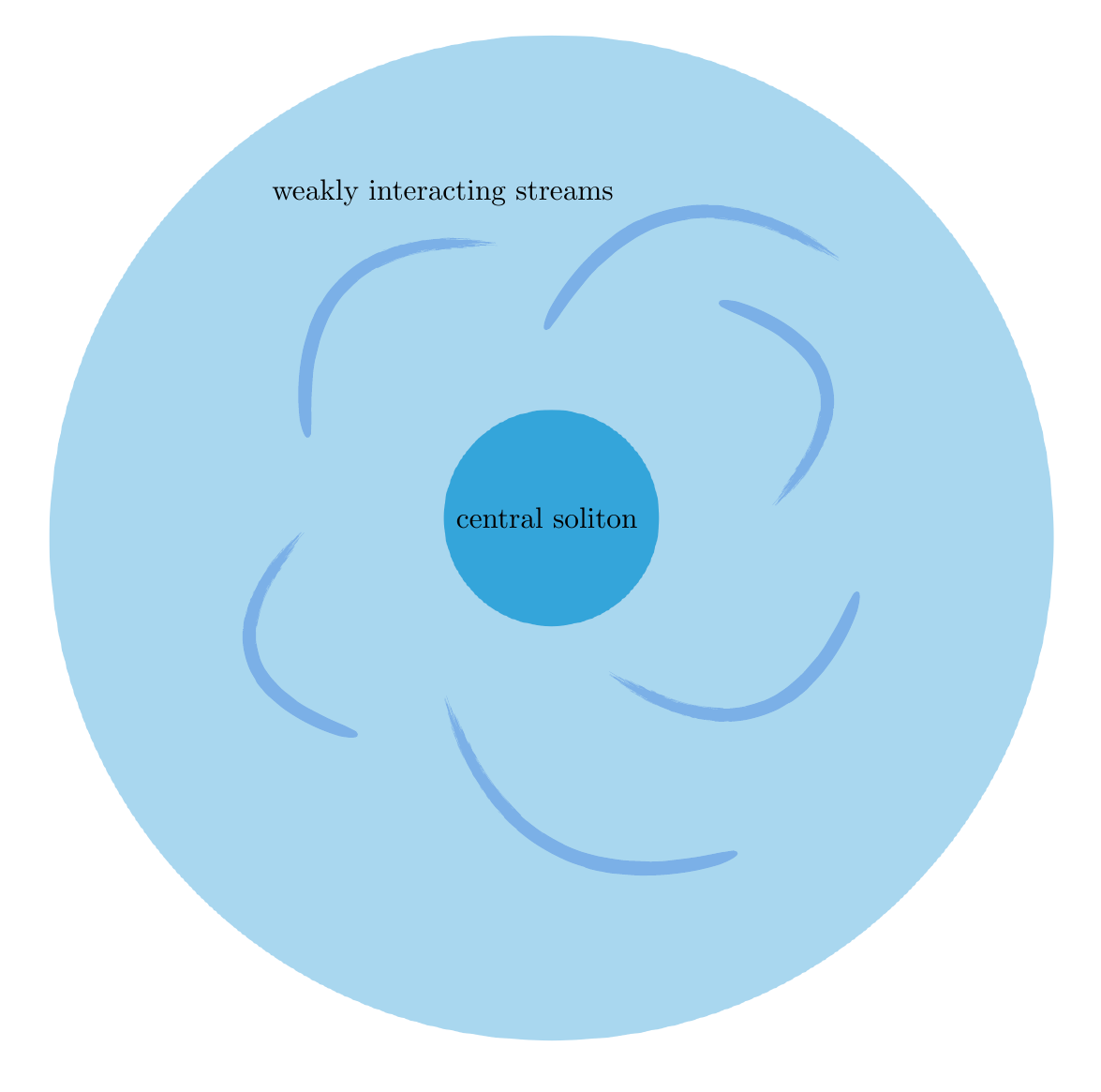}
	\caption{\it Pictorial representation of the central superfluid region of size~$R_{\rm T}$. This region includes a central superfluid soliton, surrounded by a collection of weakly interacting streams of superfluid debris orbiting around. This debris is the result of the tidal disruption of the population of solitons that formed in the outskirts of the thermal core. Depending on whether the halo is in thermal equilibrium or not, an outskirt of out-of-equilibrium degenerate particles is present outside the thermal core, up to the virial radius of the halo.}
	\label{streams}
\end{figure}

However, the collection of solitons that would reside in the halo is not expected to be stable. For solitons residing outside the central region of the halo, the assumption of self-gravitation breaks down. In~\cite{Berezhiani:2022buv}, it was shown tidal effects are effective in destroying them. The net result of this process is the formation of weakly-interacting superfluid streams, orbiting around a leftover central soliton.\footnote{Self-interacting solitons embedded in a diffuse halo are expected to emerge in models of self-interacting dark matter as studied in Ref.~\cite{Garcia:2023abs}, with size dependent on the halo formation history.} See~Fig.~\ref{streams} for a pictorial representation. Neglecting baryons, this complex distribution is then expected to match a coarse-grained NFW profile~\cite{Navarro:1995iw, Navarro:1996gj} due to the weakly-interacting nature of the debris. We refer the reader to the paper \cite{Berezhiani:2022buv} for more details on the formation of tidal streams.

Finally, we can quantify an upper bound on the size of the central soliton. In order to see how comfortable we are in the choice of the model parameters to get a superfluid soliton of tens of kpcs in size, it is useful to express the equation of state at matter-radiation equality in terms of the Jeans scale~\cite{Berezhiani:2022buv}
\be
\frac{P}{\rho}\bigg|_\text{\tiny eq} \simeq 10^{-5} \lp \frac{\lambda_{\rm J}}{\rm kpc} \rp^2\,.
\ee
One could therefore admit sizes of about 30 kpc to get a ratio of pressure and density at matter-radiation equality of~$(P/\rho)|_\text{\tiny eq} \simeq 0.01$, which is consistent with CMB observations~\cite{Avelino:2015dwa}.

\subsection{Superfluid density profile around black holes}
Galaxies are expected to host a supermassive BH at their center. The presence of a supermassive BH modifies the density profile of the central superfluid soliton, due to its gravitational potential. One expects that the density profile steepens in its vicinity, from the otherwise nearly homogeneous core, and that DM self-interactions are correspondingly enhanced. In this Section, we briefly summarize the main results derived in Ref.~\cite{DeLuca:2023laa}, where the interested reader can find additional details.

We consider a non-spinning, central black hole with mass~$M_\BH$ and radius given by
\be 
\label{Schwarzschild}
r_\BH = 2 G M_\BH \simeq 9.6 \, \cdot 10^{-11}\, {\rm kpc} \lp \frac{M_\BH}{10^6 M_\odot}\rp\,.
\ee
Note that we have in mind a supermassive BH which can potentially be observed with LISA.
The BH gravitational well is expected to modify the DM distribution within the characteristic radius 
\be
r_h = \lp \frac{3M_\BH}{4 \pi \rho_0} \rp^{1/3} 
\simeq 0.25 \, {\rm kpc} \lp \frac{M_\BH}{10^{6} M_\odot}\rp^{1/3} \lp \frac{\rho_0}{10^{-24} {\rm g/cm^3}} \rp^{-1/3}\,.
\ee
By definition,~$r_h$ is the radius at which the DM potential equals the potential induced by the BH or, equivalently, where the total enclosed DM mass and the BH mass are comparable.  To simplify the problem, we assume a  spherically-symmetric  DM distribution, which is also 
isotropic in velocity and has relaxed to a near-equilibrium state. 
We further assume that the BH mass is much smaller than the total integrated
DM halo mass~$M_\DM$, while it dominates the total mass of all particles bound to it in the cusp. 

We describe the density profile in three different regions. In the outer region, beyond the BH sphere of influence~$r_h$, the BH has negligible effects on the superfluid DM. The latter is therefore described by a superfluid core and its match to the standard NFW envelope, as discussed in the previous Section. In the intermediate region, at radii smaller than~$r_h$, the BH starts to modify the DM density profile because of its gravitational well. It continues until relativistic effects to the DM motion become so important that they modify the superfluid properties and equation of state. At this point, an inner region starts, where the DM profile is further modified. The interested reader can find a more detailed discussion in Ref.~\cite{DeLuca:2023laa}.

In the following, we will briefly review the DM profile under the influence of the BH. 
At distances~$r \lesssim r_h$, the BH starts to modify the density profile due to its potential well. Within this region, the DM particles are gravitationally bounded to the BH, with characteristic velocity
\be
v(r) = \sqrt{\frac{G M_\BH}{r}}\,.
\label{v interm}
\ee
Assuming energy equipartition, the DM particles can be described as having a temperature
\be
T (r) = \frac{1}{3} \frac{G m_\DM M_\BH}{r}\,.
\label{TBH}
\ee
Within this region, we will assume that the BH influence is, however, not strong enough to modify the equation of state of the DM fluid. The DM density profile can then be computed by solving the hydrostatic equilibrium equation
\be
\frac{1}{\rho (r)} \frac{\d P(r)}{\d r} = 
- \frac{4 \pi G}{r^2} \int^r_0 \d r' r'^2\rho(r') - \frac{G M_\BH}{r^2} \simeq - \frac{G M_\BH}{r^2}.
\label{hydrostatic 1}
\ee 
The first term of the middle expression comes from the superfluid self-gravity, while the second term
from the BH potential~$\Phi_\BH = - G M_\BH/r$. In the last step, we have made the simplifying assumption that at radii smaller than~$r_h$ the BH dominates the gravitational potential. 

At sufficiently small radii, when DM velocities approach the speed of light, the non-relativistic approximation breaks down. Neglecting the small contribution to the DM stress-energy tensor, one can adopt the Schwarzschild metric to describe the spherically-symmetric space-time as
\begin{equation}
{\rm d}s^2 = - \left(1-\frac{2G M_\BH}{r}\right) {\rm d}t^2 + \frac{{\rm d} r^2}{1-\frac{2G M_\BH}{r}} + r^2 {\rm d}\Omega^2\,,
\end{equation}
such that the relativistic generalization of Eq.~\eqref{hydrostatic 1} reads~\cite{Shapiro:2014oha}
\be
\frac{\d P(r)}{\d r} = - \frac{\rho (r)+P(r)}{1-2 G M_\BH/r} \frac{G M_\BH}{r^2}\,.
\label{hydro rel}
\ee
The solution to this equation provides the density profile of the superfluid DM in the intermediate region.

From Eq.~\eqref{TBH}, the temperature of the superfluid is connected to the distance from the black hole. Because of this, there exists a specific scale~$r_\text{\tiny deg}$~\cite{DeLuca:2023laa} at which the temperature of the system may exceed the critical temperature of the superfluid.
In this case, the condition of degeneracy for Bose-Einstein condensation breaks down, and DM is not sufficiently cold to remain in a superfluid state.
Thus the DM is no longer degenerate, and its equation of state is instead approximated by the ideal gas law because of the large number of interactions around the BH~\cite{Shapiro:2014oha}
\be
P = n T = \frac{\rho v^2}{3}\,.
\label{ideal}
\ee
In the two-body case, it turns out that the~$r_\text{\tiny deg}$ is smaller than the BH horizon, such that the DM remains in the superfluid state all the way to the BH horizon. In the three-body case, however, the degeneracy radius is larger than the BH horizon, and the evolution of the DM profile is correspondingly modified. (Note, however, that thermalization never breaks down as we approach the BH.) Following Refs.~\cite{Shapiro:2014oha, DeLuca:2023laa}, one can compute the DM density profile in this regime by solving the hydrostatic equilibrium equation together with a heat equation within the gravothermal fluid approximation~\cite{Balberg:2002ue, 1980MNRAS.191..483L, Shapiro:2014oha, Koda:2011yb}. These equations must be solved down to the radius~$r_\text{\tiny mb} = 4 G M_\BH$~\cite{Sadeghian:2013laa,Shapiro:2014oha}, which corresponds to marginally-bound circular orbits in the Schwarzschild geometry. At~$r_\text{\tiny mb} = 4 G M_\BH$, DM particles are completely accreted into the BH, and the DM density profile plummets.

\begin{figure}[t!]
	\centering
	\includegraphics[width=0.6\textwidth]{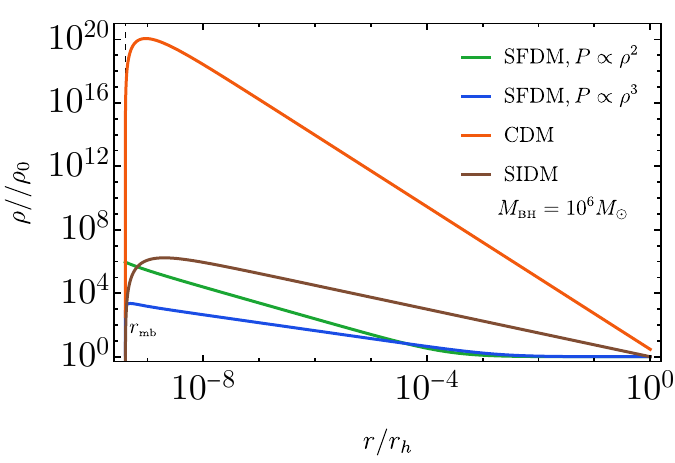}
	\caption{\it Comparison between the dark matter profile enhanced around a central BH with mass~$M_\BH = 10^6 M_\odot$ for the cases of collisionless (red), self-interacting (brown) and superfluid (green and blue) models.}
	\label{SFDMprofile}
\end{figure}

Figure~\ref{SFDMprofile} shows the density profile for the superfluid DM models discussed above, assuming a supermassive BH of mass~$M_\BH = 10^6 M_\odot$. The green and blue lines show the profile for the two- and three-body interacting case, respectively. In both cases, the profile becomes increasingly steep within the BH sphere of influence, growing by orders of magnitude compared to the case without the BH, until it drops at the accretion radius for the three-body case. 
Comparing the two interacting models, one can appreciate that the profile for~$P\propto \rho^3$ increases slightly less than in the~$P \propto \rho^2$ case. This is due to the fact that, as we go to smaller radii, the pressure in the three-body case is significantly larger, resulting in an overall milder density growth.

The role of pressure (or self-interactions) crucially impacts the growth of the density profile. This is manifest when one considers the standard case of collisionless dark matter (CDM), as studied in the pioneering work by Gondolo and Silk~\cite{Gondolo:1999ef}. From an initially isotropic profile of the form~$\rho = \rho_0 (r/r_h)^{-\gamma}$, they studied the evolution of a point mass at the center of the distribution, which grows adiabatically by accreting particles and gives rise to a massive BH. The dark matter feels the BH potential and evolves until it reaches the final configuration~\cite{Gondolo:1999ef}
\be
\label{GondoloSilkprofile}
\rho \simeq \rho_0 \lp \frac{M_\BH}{\rho_0 r_h^3} \rp^{\frac{\gamma-3}{\gamma-4}} \left(1 -\frac{r_\text{\tiny mb}}{r}\right)^3 \lp \frac{r}{r_h} \rp^{ -\frac{9-2\gamma}{4-\gamma}}\,.
\ee
In Fig.~\ref{SFDMprofile} we show the behavior of the cusp for collisionless DM assuming no tilt~$\gamma =0$ for the initial profile. As one can appreciate, the absence of self-interactions results into a much larger growth (about fifteen orders of magnitude) with respect to the case of superfluid DM. The relevance of interactions is further confirmed if one considers a model of self-interacting dark matter (SIDM), studied in Ref.~\cite{Shapiro:2014oha}. Assuming a velocity-independent cross section, it was found that the density profile has an approximate slope of~$\rho_\text{\tiny SIDM} \propto r^{-3/4}$. This model is shown in brown in the Fig.~\ref{SFDMprofile}, highlighting the overall trend that weaker interactions further enhance the profile compared to models with larger pressure, as is the case of superfluids. The huge hierarchy between these different DM models will manifest in the strength of the dynamical friction force experienced by a perturber moving through the fluids.

\section{Dynamical friction: circular orbits}
\label{Sec3}
Dynamical friction is a dissipative phenomenon experienced by a massive object in motion within a medium. It corresponds to the gradual depletion of the body's momentum due to its gravitational interactions with the wake it generates in the surrounding medium. 
Since the pioneering work of Chandrasekhar \cite{Chandrasekhar:1943ys}, the standard setup for studying dynamical friction corresponds to considering a body in linear motion with constant velocity. This setup has been widely studied for different media, such as collisional gases~\cite{Ostriker:1998fa}, superfluids~\cite{Berezhiani:2019pzd} and fuzzy dark matter~\cite{Lancaster:2019mde}.

Although the case of linear motion may provide an accurate approximation of dynamical friction for general trajectories, this approximation breaks down when the timescale over which friction significantly alters the dynamics of the massive object becomes comparable to the timescale over which the general trajectory deviates from linear motion. 
Binaries enter this last category, as the effect of dynamical friction may be important over several orbital cycles. 
In this Section, we derive the dynamical friction of bodies moving in a circular motion in a homogeneous superfluid medium, following the hydrodynamic set-up employed in Refs.~\cite{2022ApJ...928...64D,Buehler:2022tmr}. We sum up the preliminary steps and refer the reader to these papers for more details.

Consider an external probe of mass~$\mu$ (which in the next Section will stand for the binary reduced mass) moving in a homogeneous superfluid medium of density~$\rho_0$. The linear fluid response of the medium to the orbiting probe is determined by solving the Euler and continuity equations for the fluid density and velocity~\cite{Boehmer:2007um},
\begin{align}
\frac{\partial \rho}{\partial t} + \vec \nabla \cdot (\rho \vec v) &= 0\,; \nonumber \\
\frac{\partial \vec v}{ \partial t} + (\vec v \cdot \vec \nabla) \vec v &= - \frac{1}{\rho} \vec \nabla P - \vec \nabla \Phi_\DM + \frac{1}{2 m_\DM^2} \vec \nabla \left( \frac{\Delta \sqrt \rho}{\sqrt \rho} \right)\,,
\label{navier-stokes}
\end{align}
which represent the hydrodynamical form of the  nonrelativistic Gross-Pitaevskii equation,
where the last term in the second equation identifies the so-called “quantum pressure” term.
One perturbs these hydrodynamical equations around a static background of homogeneous density~$\rho_0$ as 
\be
\rho(\vec{r},t) = \rho_0\Big(1 + \alpha(\vec{r},t)\Big)\,,
\ee
where the density inhomogeneity~$\alpha$ is created by the probe moving in the fluid.
This density inhomogeneity, together with the external perturber~$\rho_\text{\tiny ext} = \mu \, \delta^{(3)}\big(\vec{r}-\vec{r}_p(t)\big)$, both source a perturbation~$\phi$ in the gravitational potential:
\be
\vec{\nabla}^2 \phi = 4 \pi G \left(\rho_0 \alpha + \mu \, \delta^{(3)}\big(\vec{r}-\vec{r}_p(t)\big)\right)\,,
\label{Poisson}
\ee
with~$\vec{r}_p(t)$ denoting the probe position at time~$t$. It is easy to show that the perturbed equations can be recast as a second-order equation for the density inhomogeneity~\cite{2022ApJ...928...64D,Buehler:2022tmr},
\begin{equation}
    \frac{\partial^2}{\partial t^2} \alpha(\vec{r},t)-c_s^2\vec{\nabla}^2 \alpha (\vec{r},t) +\frac{1}{4m_\DM^2}\vec{\nabla}^4\alpha(\vec{r},t)=4\pi G \mu \delta^{(3)}\big(\vec{r}-\vec{r}_p(t)\big)\,.
    \label{eq:eqPert}
\end{equation}
On the left-hand side, we have neglected the gravitational mass term coming from self-gravity,~$- m_g^2 \alpha = - 4 \pi G \rho_0 \alpha$~\cite{Berezhiani:2019pzd}. This term is relevant for modes with sufficiently low momenta in the dispersion relation. Its effects on dynamical friction have been described in Ref.~\cite{Berezhiani:2019pzd} for a linear trajectory. We briefly discuss its role in Sec.~\ref{quantumpressureregime}, but leave its complete inclusion to future work. Furthermore, we stress that adopting this expression for the dispersion relation, $\omega_k^2 = c_s^2 k^2 + \frac{k^4}{4 m_\DM^2}$, allows us to study also the region of momenta smaller than the chemical potential, discussed in Appendix~\ref{AppA}, with a redefinition of the effective sound speed and particle mass.

Equation~\eqref{eq:eqPert} can be solved in terms of the retarded Green function, through a convolution with the source term~\cite{2022ApJ...928...64D,Buehler:2022tmr}. Dynamical friction is then derived as the gravitational attraction between the wake~$\alpha(\vec{r},t)$ and the probe as
\begin{equation}
  \label{eq:df1}
\vec{F}_\text{\tiny DF}= (4\pi G \mu)^2 \rho_0 \int_0^\infty {\rm d}\tau \int \frac{{\rm d}^4 k}{k^2} \frac{{\rm i}\vec{k} }{c_s^2 k^2 + \frac{k^4}{4 m_\DM^2}-(\omega+{\rm i}\epsilon)^2} {\rm e}^{-{\rm i}\omega \tau+{\rm i} \vec{k}\cdot\left(\vec{r}_p-\pvec{r}'_p \right)} \,,
\end{equation}
where~$\pvec{r}'_p(t)=\vec{r}_p(t-\tau)$ is the position of the probe evaluated at the retarded time~$t-\tau$ and $\epsilon >0$ enforces causality. 
Notice that the formula \eqref{eq:df1} is general and assumes only the homogeneity of the unperturbed medium.

Specializing now to a probe moving on a circular orbit of radius~$r_0$ and constant angular velocity~$\Omega$, we parameterize~$\vec{r}_p(t)$ in spherical coordinates, choosing the motion to lie in the equatorial plane:
\begin{equation}
    \vec{r}_p (t) =\left(r_0, \, \frac{\pi}{2}, \, \Omega t\right)\,.
\end{equation}
Using the standard spherical harmonics decomposition, the exponentials can be expanded as
\begin{equation}
    {\rm e}^{{\rm i} \vec{k} \cdot \vec{r}}=4 \pi \sum_{\ell=0}^\infty\sum_{-m}^m {\rm i}^\ell j_\ell(k r) Y_\ell^m(\hat{k}) {Y_\ell^m(\hat{r})}\,.
\end{equation}
Meanwhile, in the polarization basis~$(\hat{e}_+,\hat{e}_-,\hat{z})$, with~$\hat{e}_\pm=\frac{1}{\sqrt{2}}\left({\rm i}\hat{y}\mp \hat{x}\right)$,
the momentum vector~$\vec{k}$ is decomposed as
\begin{equation}
    \vec{k}=\sqrt{\frac{4}{3}\pi}k\left(Y_1^0(\hat{k})\hat{z}+Y_1^{+1}(\hat{k})\hat{e}_++Y_1^{-1}(\hat{k})\hat{e}_-\right)\,.
\end{equation}
Using these, the friction force can be simplified to
\begin{flalign}
    \vec{F}_\DF & = {\rm i}\sqrt{\frac{4}{3}\pi}(4\pi G \mu)^2 \rho_0 \sum_{\ell_1,\ell_2,m_1,m_2} \int_0^\infty {\rm d}\tau {\rm e}^{-{\rm i}\omega \tau} \int {{\rm d}k \, k \, {\rm d}\omega \, {\rm d}\Omega} \, \frac{j_{\ell_1}(k r_0) j_{\ell_2}(k r_0) }{\omega_k^2-(\omega+{\rm i}\epsilon)^2} \nonumber \\
    &~~~~~\times Y_{\ell_1}^{m_1}(\hat{k}){Y_{\ell_1}^{m_1}}^*(\hat{r}_p)Y_{\ell_2}^{m_2}(\hat{k}){Y_{\ell_2}^{m_2}}^*(\hat{r}'_p) \left(Y_1^0(\hat{k})\hat{z}+Y_1^{+1}(\hat{k})\hat{e}_++Y_1^{-1}(\hat{k})\hat{e}_-\right)\,.
\end{flalign}
Since the circular motion lies in the plane~$x-y$, spherical symmetry implies that the force along the~$\hat{z}$ direction vanishes. Furthermore, since the drag force is a real vector, one can further simplify the integrals in the~$\hat{y}$ direction and integrate over the solid sphere using Wigner 3j symbols, such that the expression reduces to a~$1+1$ dimensional problem. 

The resulting force is given by~\cite{2022ApJ...928...64D,Buehler:2022tmr}
\be
\vec{F}_\text{\tiny DF} =- \frac{4\pi G^2 \mu^2 \rho_0}{c_s^2}\vec{\mathcal{F}}_\text{\tiny DF}\,, 
\ee
where we have introduced a dimensionless force~$\vec{\mathcal{F}}_\text{\tiny DF}$ in the form of a sum over angular multipoles~$(\ell,m)$:
\be
\vec{\mathcal{F}}_\text{\tiny DF} \equiv c_s^2 \sum_{\ell=1}^{\ell_\text{\tiny max}}\sum_{m=-\ell}^{\ell-2}\gamma_{\ell m}\left\{\text{Re}\left(S_{\ell,\ell-1}^m-{S^{m+1}_{\ell,\ell-1}}^*\right)\hat{r}+\text{Im}\left(S_{\ell,\ell-1}^m-{S^{m+1}_{\ell,\ell-1}}^*\right)\hat{\varphi}\right\}\,.
\label{FDF1}
\ee
The cutoff~$\ell_\text{\tiny max}$ will be discussed below. The unit vectors~$\hat{r}$ and~$\hat{\varphi}$ point, respectively, in the directions perpendicular and tangential to the circular orbit. Equation~\eqref{FDF1} thus provides the perpendicular and tangential components of the drag force experienced by a probe moving on circular orbits. Notice that the assumption of circular orbit is justified as long as the acceleration induced by the drag force is smaller than the orbital acceleration. We will discuss the consistency of this assumption in the next Section. 
The above expression involves two functions,~$\gamma_{\ell m}$ and~~$S_{\ell,\ell-1}^m$, discussed below.

The first function,~$\gamma_{\ell m}$, is the result of the angular integration of the original integral~\eqref{eq:df1}. It is given by
\begin{equation}
    \gamma_{\ell m}=  (-1)^m \frac{(\ell-m)!}{(\ell-m-2)!}\left\{{\Gamma\left(\frac{1-\ell-m}{2}\right)\Gamma\left(1+\frac{\ell-m}{2}\right)\Gamma\left(\frac{3-\ell+m}{2}\right)\Gamma\left(1+\frac{\ell+m}{2}\right)}\right\}^{-1}\,,
\end{equation}
in terms of Gamma functions~$\Gamma$. Notice that~$\gamma_{\ell m}$ is not affected by changing the microscopic properties of the medium, and it is completely determined by the trajectory and the assumed homogeneity of the background. 

The second function,~$S_{\ell,\ell-1}^m$, is the~$1+1$ dimensional analog of the starting integral in Eq.~\eqref{eq:df1}. For steady-state motion, it reads
\begin{flalign}
\label{IntS}
   S_{\ell,\ell-1}^m =& \int \frac{{\rm d}\tau {\rm d}\omega}{2\pi} {\rm e}^{-{\rm i}\left(\omega-m \Omega\right)} \int k {\rm d}k\frac{j_\ell(k r_0)j_{\ell-1}(k r_0)}{c_s^2 k^2+\frac{k^4}{4m_\DM^2}-(\omega+{\rm i}\epsilon)^2}\nonumber \\  =&\int k {\rm d}k\frac{j_\ell(k r_0)j_{\ell-1}(k r_0)}{c_s^2 k^2+\frac{k^4}{4m_\DM^2}-(m \Omega +{\rm i}\epsilon)^2}\,.
\end{flalign}
Thus the problem of evaluating dynamical friction boils down to evaluating~$S_{\ell,\ell-1}^m$. The next subsections are dedicated to solve the integral in different regimes of dark matter interactions and masses.

\subsection{Full result and discussion}

 In this Section, we evaluate the function~$S_{\ell,\ell-1}^m$ by applying the Cauchy integral formula. The procedure is detailed in Appendix~\ref{AppB}, and we report here the result and main steps of the computation. First of all, we notice that in the procedure of identifying the poles of Eq.~\eqref{IntS}, the following dimensionless quantities emerge.  The first quantity is the Mach number
\begin{equation}
\mathcal{M}=\frac{r_0\Omega}{c_s} = \frac{v}{c_s}\,,
\end{equation}
which characterizes whether the motion of the perturber through the medium is subsonic~$(\mathcal{M} < 1)$ or supersonic~$(\mathcal{M} > 1)$. The second quantity is
\begin{equation}
    \ell_{\rm q}= \frac{r_0 m_\DM c_s}{\mathcal{M}} = \frac{m_\DM c_s^2}{\Omega}\,.
\end{equation}
This describes the value of the azimuthal quantum number~$m$ above which the~$k^4$ term in the dispersion relation becomes important.  
Similar to the case of linear motion, the more the probe moves supersonically, the more important is the quantum pressure in characterizing the friction. 

In the process of identifying the poles,~$\ell_{\rm q}$ and~$\mathcal{M}$ enter through the following two functions
\begin{align}
f^{\pm}_m & = \sqrt{2\pm 2\sqrt{1+\left(\frac{m}{\ell_{\rm q}}\right)^2}} \quad \Longrightarrow
\quad 
\begin{cases} f_m ^+\simeq 2\,,~~ f^-_m \simeq \frac{m}{\ell_{\rm q}}\quad  ~~~~~\text{for} \quad m \ll \ell_{\rm q}\\
f^{\pm}_m\simeq \sqrt{\frac{2m}{\ell_{\rm q}}} \quad ~~~~~~~~~~~~~~\text{for} \quad m \gg \ell_{\rm q} \,.
\end{cases}
\label{fpm def}
\end{align}
With these dimensionless quantities and functions at hand, we may evaluate the integral~\eqref{IntS} as described in Appendix \ref{AppB}. The result is
\begin{align}
    S_{\ell,\ell-1}^{m}&= \frac{\pi {\rm i}}{2c_s^2 \sqrt{1+\frac{m^2}{\ell_{\rm q}^2}} }\biggr[(-1)^{1+\theta(m)}j_\ell\big(\ell_{\rm q} \mathcal{M}  f^-_m \big)j_{\ell-1}\big(\ell_{\rm q} \mathcal{M}  f^-_m \big)
    \nonumber \\
    & 
    ~~~~~~~~~~~~~~~ +~{\rm i} j_\ell\big( \ell_{\rm q} \mathcal{M}  f^-_m \big)y_{\ell-1}\big(\ell_{\rm q} \mathcal{M}  f^-_m \big) 
   +\frac{2}{\pi}i_\ell\big( \ell_{\rm q} \mathcal{M}  f^+_m\big)k_{\ell-1}\big( \ell_{\rm q} \mathcal{M}  f^+_m\big)\biggr]\,,
\label{fullFr}
\end{align}
where~$i_\ell$ and~$k_\ell$ are the modified spherical Bessel functions of the first and second kind, and $\theta (m)$ denotes the Heaviside step-function. In particular, for vanishing azimuthal number, this reduces to
\begin{equation}
\label{M0full}
    S_{\ell,\ell-1}^0=\frac{\pi}{2c_s^2\left(4\ell^2-1\right)}-\frac{{\rm i} \pi}{2c_s^2}j_\ell( 2{\rm i} \ell_{\rm q} \mathcal{M} )h_{\ell-1}^{(1)}( 2{\rm i} \ell_{\rm q} \mathcal{M} )\,.
\end{equation}
The last necessary ingredient to evaluate the drag force in Eq.~\eqref{FDF1} is the maximum multipole~$\ell_\text{\tiny max}$ up to which the sum over~$\ell$ is performed. It is determined by the size~$R_\text{\tiny probe}$ of the probe, and radius~$r_0$ of the orbit: 
\begin{equation}
\ell_\text{\tiny max} =\frac{\pi r_0}{R_\text{\tiny probe}}\,.
\label{lmax}
\end{equation}
Both radial and tangential components converge as~$\ell_\text{\tiny max} \rightarrow \infty$. For fixed Mach number, the tangential force starts converging to its asymptotic value only after the multipole moment~$\ell_{\rm q}$, which carries the information of quantum pressure, is reached. The growth in the tangential component for low multipoles as~$\ell_\text{\tiny max}$ increases is reminiscent of the Coulomb logarithm encountered in the case of a gaseous medium, to be discussed below, which is regulated only when multipoles probing the~$k^4/4m_\DM^2$ part of the dispersion relation are included. 
In other words, when~$\ell_\text{\tiny max} > \ell_{\rm q}$, quantum pressure provides the regulator of the theory and the tangential component converges, whereas if~$\ell_\text{\tiny max} < \ell_{\rm q}$, the size of the probe provides the effective cutoff.

\begin{figure}[t!]
\centering
	\includegraphics[width=0.49\textwidth]{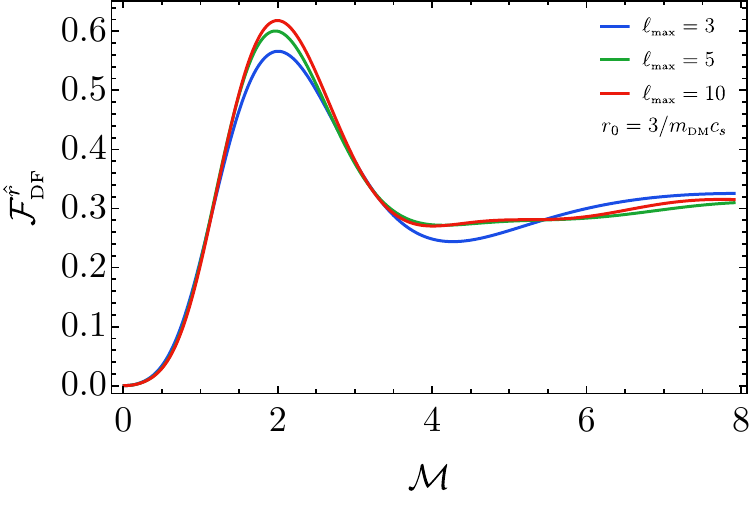}
 \includegraphics[width=0.49\textwidth]{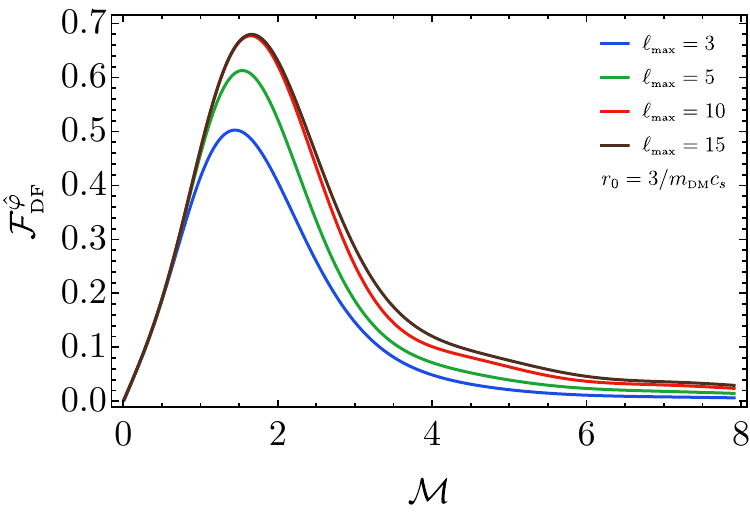}
	\caption{\it Radial (left panel) and azimuthal (right panel) components of the dynamical friction force in terms of the Mach number~$\mathcal{M}$. The orbital radius is fixed to~$r_0 = 3/m_\DM c_s$ in order to get an intermediate case ($\ell_{\rm q} \simeq \mathcal{O}(1)$) between the quantum pressure and sound regimes. The different lines correspond to different values of the maximum multipole~$\ell_\text{\tiny max}$ in the sum.}
	\label{SFforce}
\end{figure}

Figure~\ref{SFforce} shows the behavior of the dynamical friction force in terms of the Mach number~$\mathcal{M}$, for different choices of the size multipole~$\ell_\text{\tiny max}$, fixing the orbital radius to~$r_0 = 3/m_\DM c_s$ in order to get~$\ell_{\rm q} \simeq \mathcal{O}(1)$. In the left panel, one can see that the radial component of the friction force converges after summing up a few multipoles, and that is suppressed for subsonic velocities~$\mathcal{M}<1$. On the other hand, in the right panel, the tangential component of the friction is shown to grow as more multipoles are included. As discussed above, this is reminiscent of the Coulomb logarithm encountered in the sound regime~\cite{2022ApJ...928...64D} and which is regulated by introducing quantum pressure in the dispersion relation~\cite{Buehler:2022tmr}. Differently from the sound regime depicted in~\cite{2022ApJ...928...64D}, where the radial friction vanishes in the subsonic regime, neither the radial nor the tangential friction component vanishes. The reason is that quantum pressure provides a non-vanishing subsonic radial component, which is non-negligible in the case of low~$\ell_{\rm q}$.

\subsection{Sound regime}
\label{sound regime} 

In the limit of negligible quantum pressure, the superfluid dispersion relation simplifies to~$\omega_k \simeq c_s k$.  This limit recovers the standard sound result and can be obtained if
\be
\ell_{\rm q} \to \infty\,,
\label{limitSS}
\ee
keeping the Mach number~$\mathcal{M}$ fixed.
In this case the momentum integral simplifies to~\cite{2022ApJ...928...64D}
\begin{equation}
\label{gaseousS}
    S^{m}_{\ell,\ell-1}=\frac{{\rm i}\pi}{2 c_s^2}j_\ell\left(m\mathcal{M}\right)h^{(1)}_{\ell-1}\left( m\mathcal{M}\right)\,,
\end{equation}
with the case~$m = 0$ still determined by Eq.~\eqref{M0full}, albeit without the second term.

Let us comment on the limit~\eqref{limitSS}. While this is the correct limit to recover the sound regime, where all multipoles are dominated by sound modes and the dispersion relation is led by the~$c_s^2 k^2 $ term, in the case of subsonic motion the condition~$\ell_{\rm q}\gtrsim \mathcal{O}(1)$ serves as a reasonable approximation for computing the tangential drag force.\footnote{Concerning the radial drag force, both subsonic and supersonic regimes are well-approximated by including only~$\mathcal{O}(1)$ multipoles. The only inconvenience is that subsonic radial friction is exactly vanishing only if all multipoles are summed up, while spurious, yet small, contributions are obtained within this approximation.} This approximation works fairly well due to the rapid convergence of dynamical friction in the subsonic limit, once a few multipoles are summed up. 
Notice that this fast convergence is not there in the supersonic regime, due to the presence of the Coulomb logarithm. Finally, let us mention that in the intermediate case~$1 \ll \ell_{\rm q} < \ell_\text{\tiny max}$, the setup can be modelled as a sound regime with an effective cutoff given by a size~$\sim (m_\DM c_s)^{-1}$. In this regime, multipoles between~$\ell_{\rm q}$ and~$\ell_\text{\tiny max}$ are quantum-pressure dominated and do not contribute to the Coulomb logarithm.

\begin{figure}[t!]
\centering
  \includegraphics[width=0.328\textwidth]{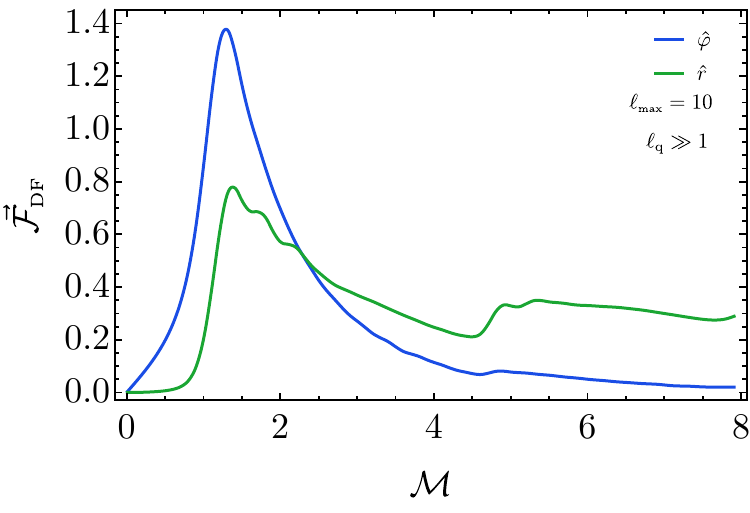}
    \includegraphics[width=0.328\textwidth]{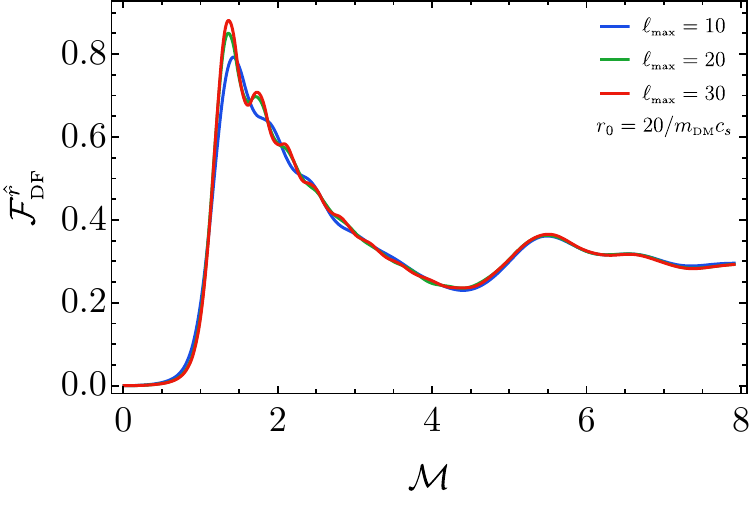}
    \includegraphics[width=0.328\textwidth]{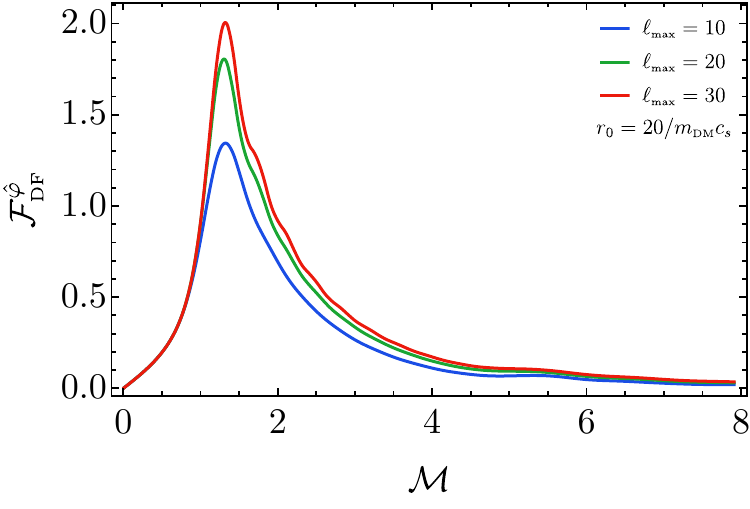}
	\caption{\it Left panel: drag force in the sound speed regime in terms of the Mach number~$\mathcal{M}$, assuming~$\ell_\text{\tiny max} = 10$~\cite{Kim:2008ab} and~$r_0 = 100/m_\DM c_s$ to enforce~$\ell_{\rm q} \gg 1$. Central and right panels: radial and azimuthal components of the force assuming the orbital radius~$r_0 = 30/m_\DM c_s~$ to describe the sound regime. The different lines correspond to different values of the maximum multipole~$\ell_\text{\tiny max}$ in the sum.}
	\label{SFforcegas}
\end{figure}

Figure~\ref{SFforcegas} shows the drag force in the sound regime. In the subsonic limit, the tangential force is always non-vanishing for~$\mathcal{M} \lesssim 1$ and larger than the radial component. The tangential component is characterized by a large bump at~$\mathcal{M} \simeq 1.4$, analogous to the linear-trajectory case discussed in Ref.~\cite{Ostriker:1998fa}.
The radial component presents instead a series of bumps that are due to the overlapping of the Mach cone and the sonic sphere, since the perturber is able to enter its own wake on a circular orbit. These results match the ones previously obtained in Refs.~\cite{Kim:2008ab,2022ApJ...928...64D}. The right panels exhibit the convergence of the force when higher multipoles are taken into account, similarly to the full case shown in Fig.~\ref{SFforce}.

In the subsonic and sound regime,~$\mathcal{M} < 1$ and~$\ell_{\rm q}\rightarrow \infty$, the tangential component of the force can be summed up in the limit of infinite~$\ell_\text{\tiny max}$, since the tangential component of the friction is convergent in the limit of subsonic velocities. The resummed version takes the exact form
\begin{equation}
\label{Fsummedup}
\vec{F}_\text{\tiny DF} \cdot \hat{\varphi} =-\frac{4\pi G^2 \mu^2\rho_0}{v^2} \left(\text{arctanh} \mathcal{M}-\mathcal{M}\right)\,.
\end{equation}
Notice that this resummation is exact up to~$\mathcal{O}\left(m \Omega/m_\DM c_s\right)$ corrections and breaks down only for~$\mathcal{M}\geq 1$. Moreover, if we expand this result for small Mach numbers, the tangential component of the friction reduces to the drag created by a linear trajectory perturber turned on at~$t = 0$~\cite{Ostriker:1998fa}, as expected.

In contrast, the radial component does not admit an analytical summed-up version. However, the numerical summation of all multipoles confirms that this term vanishes for small Mach numbers~$\mathcal{M} \ll 1$, as illustrated in the central panel of Fig.~\ref{SFforcegas}.

\subsection{Quantum pressure regime}
\label{quantumpressureregime}
The second limit we may consider of the full result \eqref{fullFr} is the one in which quantum pressure is the main contribution to the dynamics of the wake. This scenario is obtained if~$\ell_{\rm q}$ is small,
\begin{equation}
\ell_{\rm q}< 1\,.
\end{equation}
Once this condition is saturated, the Green function determining~$S_{\ell,\ell-1}^m$ is mostly dominated by quantum pressure for every~$m$.\footnote{The value~$m=0$ is determined only by the pole at~$k=0$ and it is independent of the microscopic dynamics of the background, as long as the effect of self-gravity is not reintroduced in the computation.} In this limit, dynamical friction reduces to the case of fuzzy dark matter, with~$S_{\ell,\ell-1}^m$ determined by
\begin{flalign}
S_{\ell,\ell-1}^m=\frac{{\rm i}\pi \,m_\DM}{2\, m\Omega }\biggr\{j_\ell&\left( \sqrt{2 m \ell_{\rm q} \mathcal{M}^2} \right)h^{(1)}_{\ell-1}\left(\sqrt{2 m \ell_{\rm q} \mathcal{M}^2} \right)\nonumber\\&+\frac{2}{\pi}i_\ell\left( \sqrt{2 m \ell_{\rm q} \mathcal{M}^2}\right)k_{\ell-1}\left( \sqrt{2 m \ell_{\rm q} \mathcal{M}^2}\right) \biggr\} +\mathcal{O}\left(\frac{m_\DM^2 c_s^2}{m^2 \Omega^2}\right)\,.
\label{FDMlimit}
\end{flalign}
Notice that this is the same result as obtained for fuzzy DM in Ref.~\cite{Buehler:2022tmr}, albeit with a caveat. While the exact fuzzy DM limit is infrared divergent, with the divergence localized in the~$(\ell=1,m=0)$ term, the full result \eqref{fullFr} is not. This implies that the tail of corrections is important in characterizing the divergence of this specific contribution.

We close this Section with a discussion of IR divergences. In general, there is a natural infrared cutoff for the theory, given by the Jeans scale~$\lambda_{\rm J}$, which contributes to the dispersion relation of the fluid fluctuations as shown in Eq.~\eqref{dispersion2body}. However, in the case of superfluids, there is a second emergent scale in the problem, determined by the healing length~$(m_\DM c_s)^{-1}$.  In regularizing the expanded result \eqref{FDMlimit}, a natural cutoff would be determined by the minimum length between the Jeans scale and the healing length
\begin{equation}
\lambda_\text{\tiny IR}=\text{min}\left(\lambda_{\rm J}, \frac{2\pi}{m_\DM c_s}\right)\,.
\end{equation}
It is worth noting that this cutoff is anticipated to naturally arise in the computation if self-gravity is reintroduced.
By a comparison of these scales, it is straightforward to see that the choice of the cutoff boils down to the relative hierarchy of the Jeans mass~$m_g = \sqrt{4 \pi G \rho}$ and the chemical potential~$m_\DM c_s^2$. If the first dominates, then the dynamics of all mode functions within the fluid region are mostly determined by quantum pressure. This is the case in which the full result of Eq.~\eqref{fullFr} reduces to the fuzzy dark matter limit depicted in Ref.~\cite{Buehler:2022tmr}.

On the other hand, if~$m_\DM  c_s^2\gg m_g$, the infrared cutoff is determined by the healing length. In this case, the dynamics of the fluid fluctuations on the scale~$k\sim r_0^{-1}$ is dominated by the quantum pressure, but, as the typical momentum of the fluctuation gets softer, there is an intermediate scale~$\lambda^{-1}_{\rm J}\ll \bar{k}\ll r_0^{-1}$ at which fluctuations become sound modes. In general, these modes are not important in characterizing the dynamics of the wake, as the modes relevant for dynamical friction are typically on scales of size~$r_0$. Only the~$(\ell=1,m=0)$ term of Eq.~\eqref{fullFr} is sensitive to them due to its infrared divergent behavior.


\subsection{Summary of the Section}

We summarise the result of this Section. 
The value of the parameter $\ell_{\rm q}$ allows us to understand which regime of the theory we are probing.

Large $\ell_{\rm q} \gg 1$ indicate the dominance of the sound speed term $c_s k$ in the phonon dispersion relation. As $\ell_{\rm q}$ increases, more modes with wavelengths shorter than the size of the orbit fall within the sound regime. While these modes may not be important in the subsonic regime, where modes comparable to the size of the orbit primarily determine the friction, for supersonic motion also short wavelengths contribute significantly. This is a reminiscence of the Coulomb logarithm that is encountered in the analysis of dynamical friction in the gaseous medium. 
Therefore, we distinguish the following two sub-cases:
\begin{itemize}
    \item If $\ell_{\rm q}>\ell_\text{\tiny max}$, all modes which are relevant for the problem are in the sound regime.  This corresponds to the case where quantum pressure is negligible at all scales and in which the friction is approximated by~\cite{2022ApJ...928...64D}.
    \item If $\ell_{\rm q}<\ell_\text{\tiny max}$, quantum pressure regulates the Coulomb logarithm at scales bigger than the peculiar size of the perturber. Due to the convergence of dynamical friction, multipoles which are bigger than $\ell_q$ are not important in characterizing the friction. This scenario 
can be modelled as a sound regime with a perturber of effective size $\sim( m_\text{\tiny DM} c_s)^{-1}$. 
\end{itemize}

Conversely, small values $\ell_{\rm q} \lesssim 1$ suggest a quantum-pressure dominated regime. In this case, modes with a wavelength larger than the size of the orbit provide an important contribution to the dynamical friction. 
This is because, in the limit of zero self-interactions, there is an infrared divergence localized in the $(\ell=1,m=0)$ term, which is regulated by the healing length in the case of the superfluid medium. However, if the Jeans scale of the problem is smaller than the healing length, the first provides the regulator and the friction reduces to the result of Ref.~\cite{Buehler:2022tmr}. 
Finally, intermediate values indicate regimes where both contributions effectively contribute in the dispersion relation.

\section{Evolution of black hole binaries}
\label{Sec4}
In this Section, we can finally discuss the evolution of a black hole binary system surrounded by a superfluid dark matter environment. We focus on extreme mass-ratio inspirals, {\it i.e.},  with tiny mass ratio~$m_\BH/M_\BH \ll 1$. Such systems can be modelled within BH perturbation theory by studying the quasi-adiabatic orbital motion of a point-particle with mass~$m_\BH$ (the secondary) around a much heavier BH with mass~$M_\BH$ (the primary, taken to be non-rotating). To simplify the setup, we assume that the dark matter generates an enhanced spike only around the heavier BH, as described in the previous Sections, with the orbiting lighter BH moving through the superfluid and subject to dynamical friction. As we will show, the drag force will compete with gravitational radiation to determine the binary's evolution.

\subsection{Setting the stage}
In order to understand the role of dynamical friction on the evolution of binaries of compact objects, we consider an extreme mass-ratio inspiral system of two BHs, with masses~$M_\BH = 10^6 M_\odot$ and~$m_\BH = 10 M_\odot$ as reference values, moving in the superfluid core within a binary of radius~$r$. The choice for the mass~$M_\BH$ of the central supermassive BH is motivated by the possibility of observing such systems with space-based interferometers like LISA.
Choosing smaller values would give rise to intermediate mass-ratio inspirals, for example in the mass range~$10^2 M_\odot \lesssim M_\BH \lesssim 10^3 M_\odot$, which can eventually be observed with future third-generation GW experiments, like the Einstein Telescope and Cosmic Explorer.

\begin{figure}[t!]
	\centering
	\includegraphics[width=0.328\textwidth]{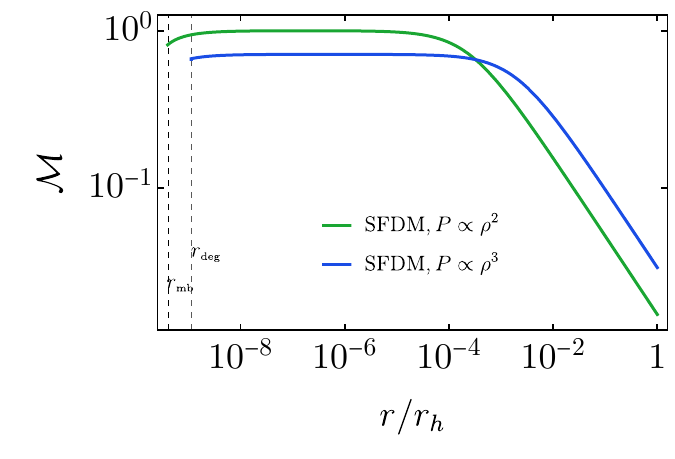}
 \includegraphics[width=0.328\textwidth]{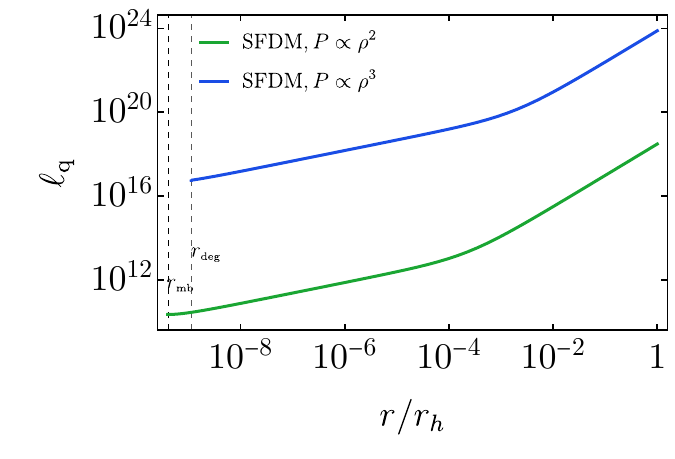}
 \includegraphics[width=0.328\textwidth]{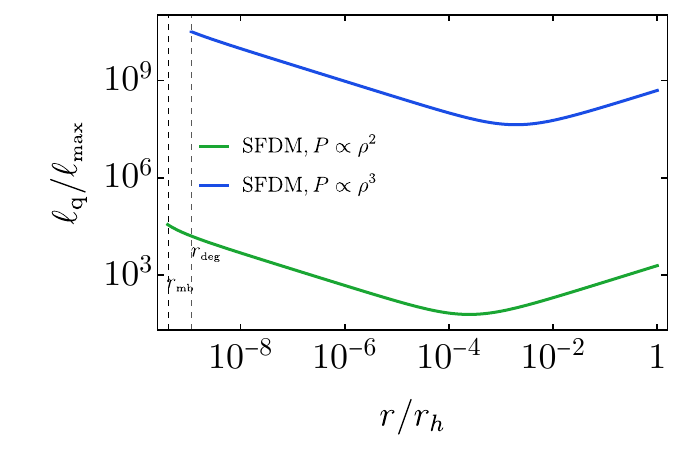}
	\caption{\it Left panel: The Mach number~$\mathcal{M} = v/c_s$ as a function of the radial distance, for the two- and three-body interacting models of superfluid dark matter. Central panel: The parameter~$\ell_{\rm q}$, which characterizes the relevance of quantum pressure, for the same models. Since~$\ell_{\rm q} \gg 1$, we are deep in the sound regime. Right panel: Comparison between~$\ell_{\rm q}$ and the finite-size cutoff~$\ell_\text{\tiny max}$. The dashed lines indicate the accretion radius~$r_\text{\tiny mb}$ and the degeneracy radius~$r_\text{\tiny deg}$ for the three-body case. We have assumed the reference values~$m_\DM = {\rm \mu eV}$ and~$m_\DM = {\rm eV}$ for the DM mass, respectively, and $\rho_0 = 10^{-25} {\rm g/cm^3}$ for the homogeneous core density.}
	\label{Stage1}
\end{figure}

As discussed in the previous Section, in order to evaluate the dynamical friction force experienced by the lighter component of the binary, let us estimate the main parameters determining the drag force. In Fig.~\ref{Stage1}, we show in the left panel the Mach number for the two and three body interacting models of superfluid DM. As one can appreciate, the evolution of the sound speed at small radii implies that we are always in the subsonic regime, reaching~$\mathcal{M} \approx 1$ only when the BH strongly modifies the DM density profile. On the other hand, before the BH starts to modify the profile around~$10^{-4} \, r_h$, we are well within the subsonic regime~$\mathcal{M} \ll 1$. In the central panel, we plot the radial evolution of the parameter~$\ell_{\rm q}$, which quantifies the relevance of quantum pressure. It is manifest that it always satisfies~$\ell_{\rm q} \gg 1$, such that within the BH sphere of influence the relevant regime for dynamical friction is the sound regime,~$\ell_{\rm q} \gg 1$, as discussed in Sec.~\ref{sound regime}.

Finally, in order to provide an analytical estimate of the drag force, we can compute the ratio of~$\ell_{\rm q}$ and the finite-size cutoff~$\ell_\text{\tiny max}$, defined in Eq.~\eqref{lmax}. The latter determines the maximum~$\ell$ in the sum over multipoles. Their ratio is given by
\begin{equation}
 \frac{\ell_{\rm q}}{\ell_\text{\tiny max}}=  \frac{m_\DM v R_\text{\tiny probe}}{\pi\mathcal{M}^2}  \simeq  \frac{10^6}{\pi\mathcal{M}^2} \lp \frac{m_\DM}{\rm eV} \rp \sqrt{\frac{r_h}{r}}\,,
\end{equation}
where in the second equality we have substituted the size of the perturber~$R_\text{\tiny probe} = 2 G m_\BH$ and the tangential velocity~$v = \Omega r  = \sqrt{G M_\BH/r}$. The estimate, supported by observing the radial behavior
  shown in the right panel of Fig.~\ref{Stage1}, proves that the ratio is always larger than unity, implying that the size provides the cutoff of the theory and that quantum pressure effects can be safely ignored. There, the reference values~$m_\DM\simeq \mu$eV (two-body case) and~$m_\DM \simeq $ eV (three-body case) have been assumed for the masses of the dark matter candidate. Let us stress that this conclusion holds only for heavy enough DM particles, in particular for the range of masses~$m_\DM \gg 10^{-9}$ eV (two-body interactions) and~$m_\DM \gg 10^{-3}$ eV (three body interactions). 
For lighter particles, it is not possible to ignore the influence of quantum pressure because, as the orbit of the binary shrinks, it would eventually reach a scale at which quantum pressure becomes the dominant contribution to the dynamics. The analysis of this general case requires a numerical investigation, since one must evolve the orbit using the complete formula for dynamical friction. We postpone this discussion to future work.

To first approximation, as long as we are in the sound regime ($\ell_{\rm q} \gg 1$) and subsonic limit (${\cal M} \ll 1$), we can focus on the tangential component of the friction force, with resummed expression given by Eq.~\eqref{Fsummedup}. An important caveat is that our derivation of dynamical friction relied on a homogeneous background, while the presence of the central BH with mass~$M_\BH$ induces a non-trivial density profile, which can in principle invalidate the computation. On the other hand, as long as the growth in density is sufficiently mild, in a sense made precise below, then an ``adiabatic" approximation is justified for dynamical friction. 

\begin{figure}[t!]
	\centering
 	\includegraphics[width=0.48\textwidth]{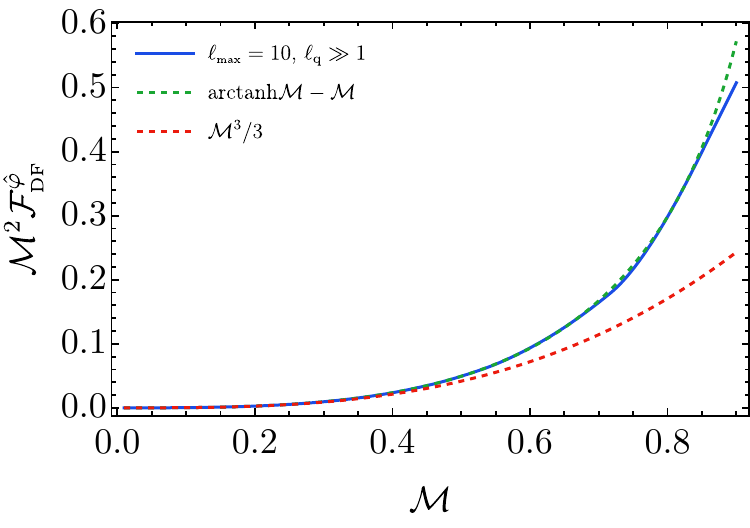}
 \includegraphics[width=0.5\textwidth]{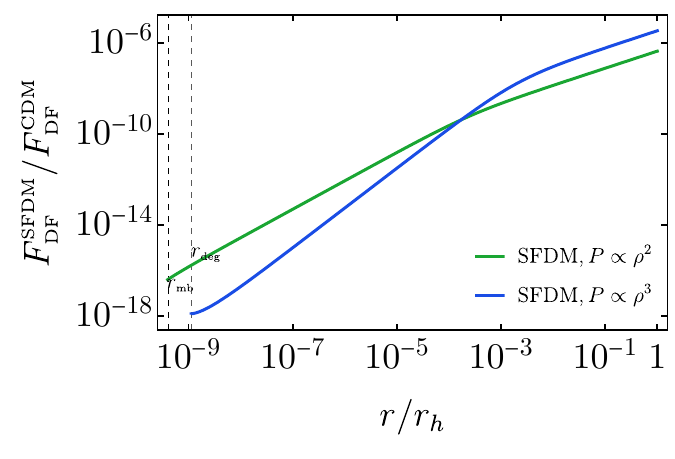}
	\caption{\it Left panel: Comparison between the numerical value of the azimuthal component of the dynamical friction force (blue line) and its analytical behavior found by resumming the multipole series in Eq.~\eqref{Fsummedup} (green line for the resummed analytical expression, and red line for its Taylor expansion). 
 Right panel: Ratio between the superfluid and collisionless friction forces as a function of the distance.}
	\label{Stage2}
\end{figure}

To see that this is indeed the case here, first note that the radial drag component for a subsonic perturber in the sound regime is suppressed.
Furthermore, along the radial direction
and over the minimum length scale relevant to estimate the drag force in the sound regime, set by the characteristic size of the overdensity~$G \mu/c_s^2$~\cite{Ostriker:1998fa}, the radial growth of the density profile is indeed negligible,
\begin{align}
 \frac{G \mu}{c_s^2} \bigg| \frac{{\rm d} \log \rho}{{\rm d} r} \bigg| \sim  \frac{G \mu}{c_s^2 r} \; \lesssim \; \lp \frac{\mu}{M_\BH} \rp \mathcal{M}^2 \ll 1\,. 
\end{align}
Thus we are justified in studying the evolution of an orbit of radius~$r$ by replacing the homogeneous density profile~$\rho_0$ in the dynamical friction formula with the non-homogeneous density profile~$\rho(r)$, evaluated at a given~$r$. In this way, Eq.~\eqref{Fsummedup} for the tangential component of the friction becomes
\begin{equation}
\label{DFforceGW}
\vec{F}_\text{\tiny DF}^\text{\tiny SFDM} \cdot \hat{\varphi} 
     \simeq -\frac{4\pi G^2 \mu^2 \rho (r)}{v^2} \frac{\mathcal{M}^3}{3}\,,
\end{equation}
where we have expanded for small Mach numbers and kept the leading order term in~${\cal M}\ll 1$. This approximate expression offers a good fit to the exact force, as shown in the left panel of Fig.~\ref{Stage2}.

For completeness, we can compare the drag force obtained for superfluids with the one for a collisionless medium, given by~\cite{Chandrasekhar:1943ys}
\be
\vec{F}_\text{\tiny DF}^\text{\tiny CDM} \cdot \hat{\varphi} 
     = -\frac{4\pi G^2 \mu^2 \rho (r)}{v^2}\,,
\ee
using the Gondolo-Silk density profile shown in Fig.~\ref{SFDMprofile}. As shown in the right panel of Fig.~\ref{Stage2}, the dynamical friction force for superfluid dark matter is orders of magnitude smaller than for collisionless dark matter. The reason for this large hierarchy is two-fold: i)~collisionless DM exhibits a stronger enhancement in density in the vicinity of the supermassive BH, as shown in Fig.~\ref{SFDMprofile}; ii)~the further suppression of the superfluid force by~${\cal M}^3$ in the subsonic regime.  This result highlights the stronger role of dynamical friction for collisionless DM than for superfluids, which will be confirmed in the next Section. Finally, comparing the superfluid drag forces in the right panel of Fig.~\ref{Stage2}, one notices that at large distances the force is larger in the three-body case because of its larger Mach number, while at smaller distances the two-body superfluid force dominates because of its further enhanced density profile.

Before discussing the role of dynamical friction in the evolution of a binary system, let us comment on the consistency of assuming a circular orbit for the motion of a perturber under the influence of a drag force. This assumption is justified as long as the change of the orbital radius in a period is smaller than the radius itself. By expressing it in terms of the ratio between the acceleration induced by the drag force and the orbital one, one gets
\begin{align}
\frac{\Delta r}{r} = \frac{|\vec{F}_\text{\tiny DF}| 4 \pi^2 r}{\mu v^2} \simeq \frac{16\pi^3}{3} \lp \frac{\mu}{M_\BH} \rp \lp \frac{\rho (r) r^3}{M_\BH} \rp \mathcal{M}^3 \ll 1\,,
\end{align}
because of the large hierarchy in the BH masses and the fact that, within $r_h$, the mass of the DM halo is much smaller than the one of the central BH.

\subsection{Gravitational wave dephasing}

After setting the stage, we can estimate the change in the trajectory of binaries,  orbiting for example in the LISA band, due to torques generated by the drag force induced by the presence of the DM environment. This change could eventually be used to detect and measure the effect of DM with GWs, as discussed for example in Refs.~\cite{Eda:2014kra, Kavanagh:2020cfn, Coogan:2021uqv}.
Following the prescription laid out in Refs.~\cite{Eda:2013gg,Eda:2014kra,Speeney:2022ryg}, one can estimate the dephasing cycles, which account for changes in the gravitational wave phase induced by dynamical friction within the DM spike. 
For simplicity, we approximate the orbit of the binary as circular, which is justified by the fact that dynamical friction circularizes inspiralling orbits within a DM spike~\cite{Becker:2021ivq}. (See, however, Ref.~\cite{Yue:2019ozq} for a discussion of its effects on eccentricity, whose increase could lead to cusps and $\mathcal{O}$(few) enhancement in the GW waveform. Nevertheless, this increase is not expected to change our main conclusions.)

The radial equation of motion for circular binaries in the Newtonian limit is given by~\cite{Speeney:2022ryg}
\begin{align}
\label{eq:radial_EoM}
\ddot{r} = -\frac{G M_\BH}{r^2} \Big(1 + \epsilon q(r)\Big) + r \Omega^2\,; \qquad  \epsilon q(r) \equiv \frac{M_\DM(r)}{M_\BH}\,,
\end{align}
where~$\Omega$ is the source's orbital frequency, and~$M_\DM(r)$ is the mass of the DM spike enclosed within a sphere of radius~$r$. Following the assumptions made in Sec.~\ref{Sec2}, one can check that the DM halo mass~$M_\DM (r) \ll M_\BH$ within~$r < r_h$ for the regions of parameter space we consider, so that the quantity~$\epsilon q(r)\ll 1$. 

Because~$\epsilon q(r)$ is small, the orbital evolution can be considered adiabatic, and the~$\ddot{r}$ term in Eq.~\eqref{eq:radial_EoM} can be neglected.
Rearranging and expanding to first order in~$\epsilon$, one obtains the following expression for~$r(\Omega)$~\cite{Speeney:2022ryg},
\begin{equation}
\label{r_of_omega}
r(\Omega) = r_0 \left ( 1 + \frac{1}{3} \epsilon q(r_0)  + \mathcal{O}(\epsilon^2) \right)\,,
\end{equation} 
where we have defined~$r_0 = (G M_\BH/\Omega^2)^{1/3}$. The adiabatic evolution of the circular orbit during the inspiral allows us to study the change in the orbital radius by computing how the orbital energy changes as the radius decreases, following the energy balance equation:
\begin{equation}
\label{energy_balance}
\dot{E}_{\text{\tiny orbit}} = \dot{E}_{\text{\tiny GW}} +\dot{E}_{\text{\tiny DF}}\,.
\end{equation}
The orbital energy loss~$\dot{E}_{\text{\tiny orbit}}$ can be expressed using the standard Newtonian expression for a circular orbit as 
\begin{equation}
\label{orbital_energy_loss}
\dot{E}_{\text{\tiny orbit}} = \frac{G M_\BH}{r^2}  \frac{\mu \dot{r}}{2} \left [ 1+\epsilon q(r) + \epsilon r q'(r) + \mathcal{O}(\epsilon^2) \right]\,,
\end{equation}
where~$\mu = m_\BH M_\BH/(m_\BH + M_\BH) \simeq m_\BH$ indicates the reduced mass of the binary and the prime denotes differentiation with respect to~$r$.
At the lowest post-Newtonian order, the GW energy loss~$\dot{E}_{\text{\tiny GW}}$ is given by the quadrupole formula, which for circular orbits reads~\cite{Maggiore:2007ulw}
\begin{equation}
\label{GW_energy_loss}
\dot{E}_{\text{\tiny GW}} = -\frac{32}{5} G \mu^2 r^4 \Omega^6\,.
\end{equation}
Lastly, the energy loss~$\dot{E}_{\text{\tiny DF}}$ due to dynamical friction is sourced as the inspiraling perturber interacts with the surrounding DM superfluid, which creates a wake behind the massive object that decelerates it. The drag force on the orbiting object is given by Eq.~\eqref{DFforceGW}, such that the energy loss becomes
\begin{equation}
\label{DF_drag_force}
\dot{E}_{\text{\tiny DF}} = \vec{v} \cdot \vec{F}_{\text{\tiny DF}} \simeq -4\pi \frac{G^2 \mu^2 \rho(r)}{v} \frac{\mathcal{M}^3}{3}\,,
\end{equation}
where only the tangential direction contributes to the evolution in the subsonic regime\footnote{Relativistic effects induced by the wake on the motion of the perturber have been neglected in the orbital evolution, see Refs.~\cite{Barausse:2007ph,Traykova:2021dua,Vicente:2022ivh} for further details.}. 

\begin{figure}[t!]
	\centering
	\includegraphics[width=0.49\textwidth]{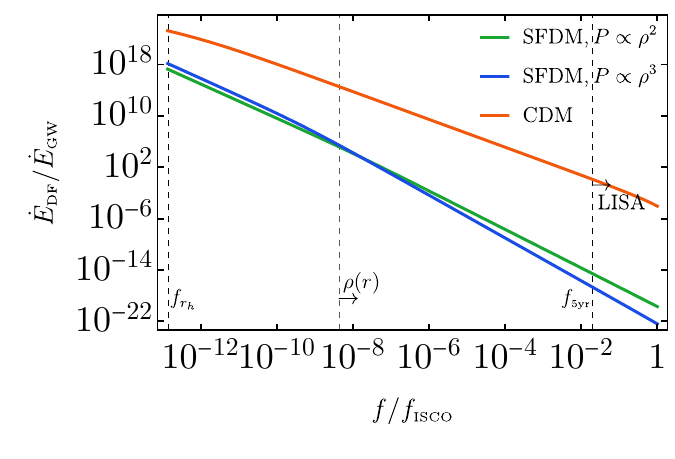}
 \includegraphics[width=0.49\textwidth]{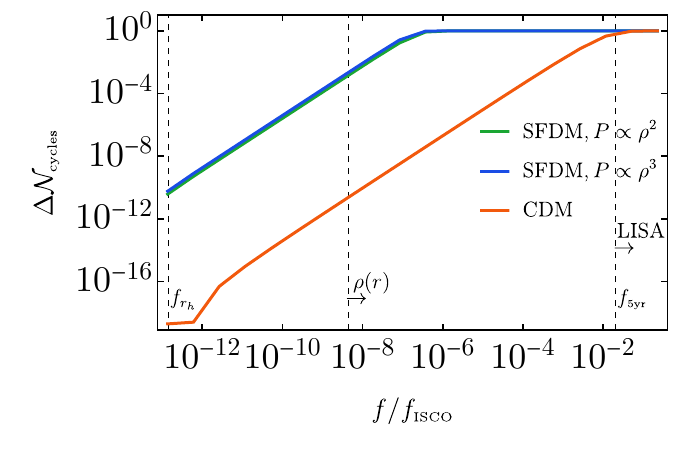}
	\caption{\it Left panel: Ratio between the dynamical friction and gravitational radiation contributions as a function of the GW frequency~$f$, for the superfluid models with two- (green) and three-body (blue) interactions, and for a collisionless/CDM medium (red line, obtained using the profile in Eq.~\eqref{GondoloSilkprofile} with~$\gamma = 0$). The vertical dashed lines indicate, respectively, the frequency~$f_{r_h}$ at which~$r = r_h$, the frequency where the BH profile starts deviating from the core,~$\rho(r) \gtrsim \rho_0$, and the frequency~$f_\text{\tiny 5yr}$ corresponding to an observation time of five years for LISA.
 Right panel: GW dephasing for the models discussed above, assuming an extreme mass-ratio inspiral with BH masses~$M_\BH = 10^6 M_\odot$ and~$m_\BH = 10 M_\odot$.}
	\label{Dephasing}
\end{figure}

The relative strength between the gravitational and drag contributions is given, at zeroth order in~$\epsilon$, by the ratio
\begin{equation}
\label{DF_correction_to_energy_loss}
\frac{\dot E_{\text{\tiny DF}}}{\dot E_{\text{\tiny GW}}} = 
\frac{5 G^{1/3} \rho}{24 \pi^{5/3} M_\BH^{2/3} f^{8/3} c_s^3}\,,
\end{equation}
in terms of the GW frequency~$f = \Omega/\pi$. This ratio is plotted in the left panel of Fig.~\ref{Dephasing} for the superfluid two- and three-body interacting cases, spanning a frequency range from the frequency at which the binary semi-major axis is equal to the BH sphere of influence,~$f_{r_h}$, to the frequency~$f_\text{\tiny ISCO}$ of the innermost stable circular orbit. The latter is computed taking into account self-force effects due to the lighter component of the binary~\cite{Favata:2010ic}, and is approximately given by
\be
f_\text{\tiny ISCO} \simeq 3 \cdot 10^{-3} {\rm Hz} \lp \frac{M_\BH}{10^{6} M_\odot} \rp^{-1}\,.
\ee 
Figure~\ref{Dephasing} shows that, at small GW frequency, the binary evolution is dominated by dynamical friction. In particular, within the superfluid core, GW emission is negligible and dynamical friction provides the main channel through which the binary shrinks.
However, once the binary semi-major axis becomes comparable to the radius at which the dark matter profile develops the spike due to the black hole influence, which takes place when~$\rho(r) \gtrsim \rho_0$ at around~$r \simeq 10^{-4} r_h$ as visualized in Fig.~\ref{SFDMprofile}, then GW emission starts to dominate the binary's inspiral. For larger frequencies, GW emission becomes the predominant driving channel leading to the binary merger.\footnote{For reference, we also mention that in astrophysical environments, only binaries as wide as at most~$\sim {\rm pc}$, corresponding to an initial GW frequency of~$\sim 10^{-9}f_\text{\tiny ISCO}$ for the system at hand, are able to merge without being disrupted by scattering with stars~\cite{Maggiore:2018sht}.}

To ascertain the role of dynamical friction in the evolution of binaries observable with LISA, which is characterized by an observation time of at most five years, we have indicated by a dashed line in Fig.~\ref{Dephasing} the minimal frequency~$f_\text{\tiny 5yr}$ for which binary inspirals lie within this observation time. As one can appreciate, the relative strength between gravitational and drag contributions satisfies~$\dot E_{\text{\tiny DF}}/\dot E_{\text{\tiny GW}} \ll 1$ in the frequency range~$(f_\text{\tiny 5yr} , f_\text{\tiny ISCO})$, showing that dynamical friction for superfluid dark matter is not sizeable enough to leave an imprint in the binary's evolution.

Using Eq.~\eqref{r_of_omega}, we can invert Eq.~\eqref{energy_balance} to get the frequency evolution equation
\begin{equation}
\label{fsdot}
\dot{f} = \dot{f}^{(0)} + \epsilon \dot{f}^{(1)} + \mathcal{O}(\epsilon^2)\,,
\end{equation}
in terms of the functions 
 \begin{align}
 \label{f_s_definitions}
 \dot{f}^{(0)}&= \frac{4 \pi  f G^2 \mu  \rho }{c_s^3}
 +\frac{96 }{5} \mu\left(\pi ^{8} f^{11} G^{5}  M_\BH^{2}\right)^{1/3}
 \,;
 \\
 \dot{f}^{(1)}&= -  \mu q' \left\{\frac{128}{5} \pi^{2} f^{3} G^{2} M_\BH  + \frac{\rho}{3 c_s^3}\left( \pi^{16} {f G^{7} M_\BH \rho}\right)^{1/3}\right\}- \frac{64}{5} \mu q\,\left(\pi ^{8} f^{11} G^{5}  M_\BH^{2}\right)^{1/3} \,,\nonumber
 \end{align}
where one can easily recognize the contributions from GW emission and dynamical friction by checking the sound-speed dependence of a specific contribution.

From Eq.~\eqref{fsdot}, one can find the orbital phase in the stationary phase approximation, from which the number of GW cycles~$\mathcal{N}_\text{\tiny cycles}(f)$ can be obtained as
\begin{equation}
\label{eq:Ncycles_def}
\mathcal{N}_\text{\tiny cycles}(f) = \int_f^{f_\text{\tiny ISCO}}\frac{f'}{\dot{f}'}{\rm d}f'\,. 
\end{equation}
One can then split the contributions from dynamical friction and GW evolution by computing the GW dephasing 
\begin{equation}
\Delta \mathcal{N}_\text{\tiny cycles} = \frac{\mathcal{N}_\text{\tiny cycles}|_\text{\tiny GW + DF}}{\mathcal{N}_\text{\tiny cycles}|_\text{\tiny GW}}\,,
\end{equation}
whose behavior is shown in the right panel of Fig.~\ref{Dephasing}. Values smaller than unity,~$\Delta \mathcal{N}_\text{\tiny cycles} \lesssim 1$, indicate an efficient role of dynamical friction in driving the binary's inspiral. Similarly to what was discussed for the energy ratio~$\dot E_{\text{\tiny DF}}/\dot E_{\text{\tiny GW}}$, one can appreciate that the GW dephasing is manifest only for wide enough binaries, where dynamical friction represents the main force dictating the binary's evolution. This conclusion is robust for both models of superfluid dark matter assumed in the paper, irrespective of their interactions. Moreover, in agreement with Fig.~\ref{Dephasing}, in the range of frequencies that can be probed by LISA, the role of dynamical friction on the binaries is not large enough to be detected.

The results obtained for the GW dephasing are based on a series of assumptions in the derivation of Eq.~\eqref{DFforceGW} for the drag force. Let us briefly comment on the consistency of our approach. First, our derivation assumes subsonic motion for the lighter component in the binary. This is justified at distances~$r \gtrsim 10^{-4} \, r_h$ (see Fig.~\ref{Stage1}), where dynamical friction dominates the evolution, but it breaks down 
at smaller distances, where the Mach number reaches values~$\mathcal{O}(1)$. Even though the force may receive corrections, we do not expect our conclusions to dramatically change, both for the good agreement of the analytical expression as shown in the left panel of Fig.~\ref{Stage2}, and because at smaller distances GW emission already provides the dominant channel in driving the binary inspiral.

Second, our derivation is based on the phonon dispersion relation discussed in Appendix \ref{AppA} and shown in Eq.~\eqref{dispersion2body}, which assumes nonrelativistic DM particles and small sound speed~$c_s \ll 1$. These assumptions, however, break down at small radii, where the DM becomes relativistic (see Ref.~\cite{Bernstein:1990kf} for a detailed derivation of the phonon dispersion relation for a relativistic Bose gas and Appendix \ref{AppA} for few details). This occurs approximately at the orbital frequency
\be
f_\text{\tiny rel} = 6.4 \cdot 10^{-5} {\rm Hz} \lp \frac{M_\BH}{10^{6} M_\odot} \rp^{-1} \lp \frac{v}{0.1} \rp^{3}\,.
\ee
While a dedicated computation is needed to have a clear picture of the relativistic case, which may eventually change the details of the drag force expression, we do not expect our main conclusions to be affected, since in the regime~$f \gtrsim f_\text{\tiny rel} \simeq 10^{-2} f_\text{\tiny ISCO}$ dynamical friction is already negligible compared to GW emission.

Finally, following the comparison and discussion of Fig.~\ref{Stage2}, one can contrast these results with the ones for the case of collisionless dark matter (also discussed in Ref.~\cite{Speeney:2022ryg}, where relativistic effects are taken into account). By looking at the red lines in Fig.~\ref{Dephasing}, one can appreciate that dynamical friction for a collisionless rather than superfluid medium has a stronger role in driving the binary inspiral, also in the frequency range probed by LISA. The reason behind this difference lies in the enhancement of the density profile for the two models, see Fig.~\ref{SFDMprofile}. In particular, as already discussed above, the different nature of the interactions in the superfluid model compared to a collisionless medium is responsible for the suppressed growth of the profile as the BH horizon is approached. The large difference in the enhancement translates into a much weaker dynamical friction force for superfluids rather than collisionless dark matter.

\section{Conclusions}
\label{Sec5}
The theory of superfluid dark matter, based on the condensation and thermalization of self-interacting, sub-eV bosonic particles at the center of galaxies, provides a novel way to conciliate the triumph of the~$\Lambda$CDM model on cosmological scales with the dynamics on galactic scales. Within this theory, the dark matter bosons are able to generate a superfluid core in the galactic center, surrounded by particles in the normal phase in the outskirts of the halo.
The presence of massive black holes in these environments is however expected to modify the superfluid core, creating overdensity regions known as dark matter spikes. 

In this work, we investigated how superfluid dark matter spikes affect the evolution of binary systems orbiting in their environment by quantifying the role that dynamical friction plays on the system's inspiral. In particular, we computed the drag force experienced by a perturber assembled in an extreme mass ratio inspiral with the central supermassive black hole and compared its contribution to the binary evolution with the emission of gravitational waves. We found that the amount of 
gravitational wave dephasing for a superfluid environment is not large enough to be detected with space-based experiments like LISA. This result can be used to disentangle the superfluid theory from the standard predictions of collisionless dark matter. 

Our work provides a preliminary step to a full modelling of the evolution of a binary system within a dark matter superfluid and can be extended by including several additional effects. First of all, a dark matter spike induces a change in the mass enclosed within the orbit of the secondary body as the system inspirals, modifying the Keplerian frequency at a given orbital radius from that of a black hole in vacuum~\cite{Eda:2013gg}. The second (relativistic) effect is related to the amount of dark matter particles accreted by the lighter black hole as the binary evolves, which changes its mass and therefore impacts on the system's inspiral~\cite{Yue:2017iwc, Boudon:2022dxi,Boudon:2023qbu}. The efficiencies of accretion and dynamical friction are finally impacted by a third effect, that is dark matter feedback, based on the fact that the dark matter distribution should be jointly evolved to determine consistently its effect on the emitted gravitational waves~\cite{Kavanagh:2020cfn}. For dynamical friction this amounts to determining the local density of dark matter particles around the secondary~\cite{Kavanagh:2020cfn,Coogan:2021uqv}, while for accretion it consists of taking into account the loss of particles from the distribution function which are accreted by the secondary~\cite{Nichols:2023ufs,Boudon:2022dxi,Boudon:2023qbu}. Including these effects would represent a huge step towards a full characterisation of the evolution of a binary system in superfluids.

Furthermore, we stress that one could generalize our results to other gravitational wave experiments like Einstein Telescope and Cosmic Explorer, which would be sensitive to lighter central black holes, including as well the information of black hole spins, and study different regimes of interactions for superfluid dark matter, for example described by alternative fluid equations of state. 

Finally, the suppression of dynamical friction for superfluids  compared to cold dark matter may offer a natural explanation to the puzzle concerning the absence of mergers for the five globular clusters orbiting Fornax~\cite{1976ApJ...203..345T}, which should have long ago spiraled to the center if the dwarf galaxy is composed of a cuspy cold dark matter profile. Superfluid dark matter, characterized by longer timescales for dynamical friction, may offer a solution to understand the survival of these clusters~\cite{Berezhiani:2019pzd} (see Refs.~\cite{Hui:2016ltb, Lancaster:2019mde, Hartman:2020fbg, Bar:2021jff} for alternative models).
Moreover, superfluid dark matter could be accompanied by the formation of vortices~\cite{Berezhiani:2015bqa}, which could be stable when a central perturber (like a SMBH) is present~\cite{Dmitriev:2021utv, Glennon:2023oqa}. Their stability could lead to the transfer of angular momentum from the binary system to the fluid, therefore affecting its orbital evolution, and modify the dark matter dynamics around the compact objects.
We leave these further refinements and prospects to future work.

\subsubsection*{Acknowledgments}
V.DL. is supported by funds provided by the Center for Particle Cosmology at the University of Pennsylvania. 
The work of J.K. is supported in part by the DOE (HEP) Award DE-SC0013528.

\begin{appendices}

\section{Superfluid dispersion relation for generic interactions}
\renewcommand{\theequation}{A.\arabic{equation}}
\setcounter{equation}{0}
\label{AppA}

In this Appendix, we demonstrate that the correction term~$\frac{k^4}{4m_\DM^2}$ in the dispersion relation of the phonon spectrum, given in Eq.~\eqref{dispersion2body}, is independent of the precise form of the self-interacting potential. Indeed, it is solely determined by the fact that we assume contact interactions. This allows us to generalize the spectrum of perturbations that we use for the quartic interactions superfluid to different kinds of potentials. 

Consider the general, two-derivative~U(1)-invariant Lagrangian
\begin{equation}
\mathcal{L}=-|\partial\phi|^2-m^2|\phi|^2-V\left(|\phi|^2\right)\,,
\label{LagrGenpotential}
\end{equation}
whose main assumption is that the potential is only a function of~$|\phi|^2$. If this were not the case, for instance by allowing a  non-local structure, then different corrections to the spectrum would be obtained. This is the case of superfluid helium for example, where maxons and rotons interpolate between the linear and quadratic part of the spectrum.

One can then parameterize the superfluid in terms of the following scalar field configuration 
\begin{equation}
\phi=\left(v+{h}(\vec{x},t)\right){\rm e}^{i\mu t+i{\pi}(\vec{x},t)},\qquad \text{with}\qquad 
\mu^2=m^2+V'(v^2)\,,
\label{background}
\end{equation}
as a function of the scalar field vacuum expectation value~$v$.
Here, $V'$ is the derivative of the potential with respect of $|\phi|^2$.
Then, we plug the decomposition back into the Lagrangian of Eq.~\eqref{LagrGenpotential}. By expanding in terms of the fluctuations~${h}(\vec{x},t)$ and~${\pi}(\vec{x},t)$, we may recast the Lagrangian into the following quadratic form 
\begin{flalign}
\mathcal{L}=&-\big(\vec{\nabla} h\big)^2- v^2\big(\vec{\nabla}\pi\big)^2+4 m v h \dot{\pi}-2 v^2 h^2 V''(v^2) +\ldots
\,,
\label{Lagrah}
\end{flalign}
where we have applied the definition of chemical potential~$\mu$ (not to be confused with the binary reduced mass) and took the non-relativistic limit.

We may already guess from Eq.~\eqref{Lagrah} that the phonon dispersion relation is expected to match the one obtained for quartic self-interactions, just with a different effective sound speed. This can be checked explicitly by integrating out the radial field~${h}(x,t)$ at first order in perturbation theory, with the result
\begin{equation}
\mathcal{L} = \frac{1}{2}\dot{\pi}^2+\frac{1}{2}\pi\left(c_s^2\Delta-\frac{\Delta^2}{4 m^2}\right) \pi + \dots\,,
\end{equation}
where we have kept only quadratic terms~\footnote{Here, we applied the canonical rescaling
\begin{equation}
    \pi\rightarrow \frac{\sqrt{-\Delta+4m^2 c_s^2}}{2\sqrt{2}v m}\,\pi.
\end{equation}}.
Here we have defined the sound speed of the condensate as
\begin{equation}
\label{cs}
c_s^2=\frac{V''(v^2)}{2 m^2}
\,.
\end{equation}
This shows the consistency of our approach in using the same dispersion relation for different superfluid equations of state.

Finally, let us underline that this approach is strictly valid only in the nonrelativistic regime. For a relativistic Bose gas, the phonon dispersion relation changes and has been computed to be~\cite{Bernstein:1990kf, Babichev:2018twg}~\footnote{Notice that Refs.~\cite{Bernstein:1990kf, Babichev:2018twg} express the relativistic dispersion relation in terms of the relativistic sound speed $\overline{c}_s$, which is defined as the low momentum limit of the second derivative of the quadratic dispersion relation $\frac{1}{2}\omega^2_k$.  In this paper instead, we parametrize the relativistic dispersion relation in terms of the nonrelativistic sound speed $c_s$, defined in Eq.~\eqref{cs} and related to the relativistic one by Eq.~\eqref{effcs}.}
\be
\omega_k^2 = 2 \mu^2 (1 + c_s^2) + k^2 - 2 \mu \sqrt{k^2 + (1 + c_s^2)^2 \mu^2}\,,
\ee
which, in the small momentum limit~$k \ll \mu$, simplifies to
\be
\omega_k^2 =\frac{c_s^2}{1+c_s^2} k^2 + \frac{k^4}{4 \mu^2 (1 + c_s^2)^3} + \mathcal{O}\lp \frac{k^6}{\mu^4} \rp\,.
\ee
This expression recovers the nonrelativistic result of Eq.~\eqref{dispersion2body} for~$c_s \ll 1$, and can be rewritten in the more general form
\be
\omega_k^2 = \overline{c}_s^2 k^2 + \frac{k^4}{4 \overline{m}^2}\,,
\ee
in terms of an effective inertial mass~$\overline{m}$ and sound speed~$\overline{c}_s$ given by
\be
\label{effcs}
\overline{c}_s^2 \equiv \frac{c_s^2}{1+c_s^2}\,;\qquad \overline{m}^2 \equiv \mu^2 (1 + c_s^2)^3\,.
\ee
Its momentum dependence allows us to generalize the results obtained in Sec.~\ref{Sec3} to a different regime of DM mass and sound speed. Finally, let us notice that this expression simplifies to the standard kinetic energy of a relativistic massive particle in the limit~$c_s \to 0$.

\section{Evaluating the~$S^m_{\ell,\ell-1}$ function}
\renewcommand{\theequation}{B.\arabic{equation}}
\setcounter{equation}{0}
\label{AppB}

In this Appendix we calculate the function~$S^m_{\ell,\ell-1}$, introduced in Eq.~\eqref{IntS}, by evaluating its defining momentum integral
\begin{flalign}
   S_{\ell,\ell-1}^m =\int_0^\infty k {\rm d}k\frac{j_\ell(k r_0)j_{\ell-1}(k r_0)}{c_s^2 k^2+\frac{k^4}{4m_\DM^2}-(m \Omega+{\rm i}\epsilon)^2}\,.
\end{flalign}
The first step consists of extending the momentum integration range down to minus infinity by exploiting the parity of the integrand. Second, we utilize the relation~$j_\ell(x) = \frac{1}{2}(h^{(1)}_\ell(x) + h^{(2)}_\ell(x))$, where~$h^{(1,2)}$ are the spherical Hankel functions. In this way, we can exploit the exponential behavior
\begin{equation}
    h^{(1)}(x)\sim {\rm e}^{{\rm i} x},\qquad h^{(2)}(x)\sim {\rm e}^{-{\rm i} x}
\end{equation}
and apply the Cauchy integral formula. One can identify the poles in the integrand of~Eq.~\eqref{IntS} as
\begin{equation}
    k_0=0\,;\qquad k_{1,2}= \pm {\rm i} m_\DM c_sf_m^+\,;\qquad k_{3,4}=\pm m_\DM c_sf_m^-\,,
\end{equation}
where the functions~$f_m^\pm$ are defined in Eq.~\eqref{fpm def}. 
The corresponding residues are given by
\begin{flalign}
&\text{Res}\left(\frac{kh_\ell^{(i)}h_{\ell-1}^{(j)}}{c_s^2 k^2+\frac{k^4}{4m_\DM^2} -(m\Omega+{\rm i}\epsilon)^2},k_0\right)=(-1)^{i+1}\frac{\rm i}{m^2\Omega^2 r_0^2}\,; \nonumber \\
&\text{Res}\left(\frac{k h_\ell^{(i)}h_{\ell-1}^{(j)}}{c_s^2 k^2+\frac{k^4}{4m_\DM^2} -(m\Omega+{\rm i}\epsilon)^2},k_1\right)=-\frac{1}{c_s^2 (f_m^{+\, 2}-2)} h_\ell^{({i})}\big({\rm i} \ell_{\rm q} \mathcal{M}  f^+_m \big)h_{\ell-1}^{({j})}\big({\rm i} \ell_{\rm q} \mathcal{M}   \big)\,; \nonumber \\
&\text{Res}\left(\frac{k h_\ell^{(i)}h_{\ell-1}^{(j)}}{c_s^2 k^2+\frac{k^4}{4m_\DM^2} -(m\Omega+{\rm i}\epsilon)^2},k_2\right)=\frac{1}{c_s^2 (f_m^{+\, 2}-2)} h_\ell^{({\bar{i}})}\left({\rm i} \ell_{\rm q} \mathcal{M}  f^+_m \right)h_{\ell-1}^{({\bar{j}})}\left({\rm i} \ell_{\rm q} \mathcal{M}  f^+_m \right)\,; \nonumber \\
&\text{Res}\left(\frac{k h_\ell^{(i)}h_{\ell-1}^{(j)}}{c_s^2 k^2+\frac{k^4}{4m_\DM^2} -(m\Omega+{\rm i}\epsilon)^2},k_3\right)=\frac{1}{c_s^2 (f_m^{-\, 2}+2)} h_\ell^{(i)}\left(\ell_{\rm q} \mathcal{M} f^-_m \right)h_{\ell-1}^{(j)}\left(\ell_{\rm q} \mathcal{M}  f^-_m \right)\,; \nonumber \\
&\text{Res}\left(\frac{k h_\ell^{(i)}h_{\ell-1}^{(j)}}{c_s^2 k^2+\frac{k^4}{4m_\DM^2} -(m\Omega+{\rm i}\epsilon)^2},k_4\right)=-\frac{1}{c_s^2 (f_m^{- \, 2}+2)} h_\ell^{(\bar{i})}\left(\ell_{\rm q} \mathcal{M}  f^-_m\right)h_{\ell-1}^{(\bar{j})}\left(\ell_{\rm q} \mathcal{M}  f^-_m \right)\,,
\end{flalign}
where the indices~$i$ and~$j$ run over 1 and 2,  identifying the first and second-type spherical Hankel functions. The bar on the indices means to exchange a given Hankel function with the orthogonal one. Explicitly,~$h_\ell^{(\bar{1})} = h_\ell^{(2)}$ and~$h_\ell^{(\bar{2})} = h_\ell^{(1)}$, following the same notation of Ref.~\cite{Buehler:2022tmr}.

Contours in the complex plane are chosen depending on the value of~$i$ and~$j$: if~$\{i,j\} = \{1,1\}$, we close the contour in the upper half of the complex plane; whereas if~$\{i,j\}= \{1,2\},\{2,1\}$, and~$\{2,2\}$, we close the contour in the lower half plane. To simplify the computation, we push the~$k_0$ divergence on the positive part of the complex plane, so that we have to include it only for~$\{i,j\} = \{1,1\}$. In the following, we provide the solution of the integral in the regimes~$m>0$ and~$m<0$.

\begin{itemize}
    \item
If~$m>0$, the poles~$k_0$,~$k_1$ and~$k_3$ lie in the upper-half plane, while~$k_2$ and~$k_4$ lie in the lower-half. 
Summing up the residues, we have
\begin{align}
\nonumber 
S^{m>0}_{\ell,\ell-1} &= \frac{\pi {\rm i}}{4c_s^2 }\Bigg\{\frac{{\rm i}c_s^2}{m^2 \Omega^2 r_0^2} \\
&+ \frac{1}{2\sqrt{1+\frac{m^2}{\ell_{\rm q}^2}}}\bigg[2h_\ell^{(1)}\big( \ell_{\rm q} \mathcal{M}  f^-_m  \big)h_{\ell-1}^{(1)}\big( \ell_{\rm q} \mathcal{M} f^-_m \big)+ \left(h_\ell^{(1)}\big( \ell_{\rm q} \mathcal{M}  f^-_m  \big)h_{\ell-1}^{(2)}\big( \ell_{\rm q} \mathcal{M}  f^-_m \big)+(1\leftrightarrow 2)\right)
    \nonumber \\
&   ~~~~~~~~~~-2 h_\ell^{(1)}\big(  \ell_{\rm q} \mathcal{M}  f^+_m \big)h_{\ell-1}^{(1)}\big( {\rm i} \ell_{\rm q} \mathcal{M}  f^+_m  \big)-\left(h_\ell^{(1)}\big( {\rm i} \ell_{\rm q} \mathcal{M}  f^+_m  \big)h_{\ell-1}^{(2)}\big( {\rm i} \ell_{\rm q} \mathcal{M}  f^+_m \big)+(1\leftrightarrow 2)\right) \bigg]\Bigg\}\,. \nonumber \\
\end{align}
Using the relation for Hankel functions,
\begin{equation}
2h^{(1)}_\ell(z)h^{(1)}_{\ell-1}(z)+h^{(1)}_\ell(z)h^{(2)}_{\ell-1}(z)+h^{(2)}_\ell(z)h^{(1)}_{\ell-1}(z)=4j_\ell(z)h_{\ell-1}^{(1)}(z)-\frac{2{\rm i}}{z^2}\,,
\end{equation}
the above result reduces to
\begin{flalign}
S_{\ell,\ell-1}^{m>0} =\frac{\pi {\rm i}}{2c_s^2 \sqrt{1+\frac{m^2}{\ell_{\rm q}^2}} }\biggr[j_\ell\big(  \ell_{\rm q} \mathcal{M}  f^-_m \big)h_{\ell-1}^{(1)}\big( \ell_{\rm q} \mathcal{M}  f^-_m \big)-j_\ell\big( {\rm i} \ell_{\rm q} \mathcal{M}  f^+_m \big)h_{\ell-1}^{(1)}\big( {\rm i} \ell_{\rm q} \mathcal{M}  f^+_m \big)\biggr]\,.
\label{Sm>0fin}
\end{flalign}

\item If~$m<0$, the prescription of the Green function flips the sign, and the retarded Green function becomes an advanced one. In this case, the poles~$k_0$,~$k_1$ and~$k_4$ lie in the upper-half plane, while~$k_2$ and~$k_3$ are in the lower half. We find
\begin{align}
\nonumber 
S^{m<0}_{\ell,\ell-1} &= \frac{\pi {\rm i}}{4c_s^2 }\Bigg\{\frac{{\rm i}c_s^2}{m^2 \Omega^2 r_0^2} \\
&- \frac{1}{2\sqrt{1+\frac{m^2}{\ell_{\rm q}^2}}}\bigg[ 2h_\ell^{(2)}\big( \ell_{\rm q} \mathcal{M}  f^-_m  \big)h_{\ell-1}^{(2)}\big( \ell_{\rm q} \mathcal{M} f^-_m \big) + \left(h_\ell^{(1)}\big( \ell_{\rm q} \mathcal{M}  f^-_m  \big)h_{\ell-1}^{(2)}\big( \ell_{\rm q} \mathcal{M}  f^-_m \big) +  (1\leftrightarrow 2)\right)
    \nonumber \\
&   ~~~~~~~~~~ + 2 h_\ell^{(1)}\big(  \ell_{\rm q} \mathcal{M}  f^+_m \big)h_{\ell-1}^{(1)}\big( {\rm i} \ell_{\rm q} \mathcal{M}  f^+_m  \big) + \left(h_\ell^{(1)}\big( {\rm i} \ell_{\rm q} \mathcal{M}  f^+_m  \big)h_{\ell-1}^{(2)}\big( {\rm i} \ell_{\rm q} \mathcal{M}  f^+_m \big)+(1\leftrightarrow 2)\right) \bigg]\Bigg\}\,. \nonumber \\
\label{Sm<0}
\end{align}
In this case, the Hankel relation of interest is
\begin{equation}
2h^{(2)}_\ell(z)h^{(2)}_{\ell-1}(z)+h^{(1)}_\ell(z)h^{(2)}_{\ell-1}(z)+h^{(2)}_\ell(z)h^{(1)}_{\ell-1}(z)=4j_\ell(z)h_{\ell-1}^{(2)}(z)+\frac{2{\rm i}}{z^2}\,,
\end{equation}
such that Eq.~\eqref{Sm<0} becomes
\begin{flalign}
S_{\ell,\ell-1}^{m<0} =\frac{\pi {\rm i}}{2c_s^2 \sqrt{1+\frac{m^2}{\ell_{\rm q}^2}} }\biggr[-j_\ell\big(  \ell_{\rm q} \mathcal{M}  f^-_m \big)h_{\ell-1}^{(2)}\big( \ell_{\rm q} \mathcal{M}  f^-_m \big)-j_\ell\big( {\rm i} \ell_{\rm q} \mathcal{M}  f^+_m \big)h_{\ell-1}^{(1)}\big( {\rm i} \ell_{\rm q} \mathcal{M}  f^+_m \big)\biggr]\,.
\label{Sm<0fin}
\end{flalign}
\end{itemize}

Summing up the two contributions, given in Eqs.~\eqref{Sm>0fin} and~\eqref{Sm<0fin}, we arrive at the final formula
\begin{align}
    S_{\ell,\ell-1}^{m}&= \frac{\pi {\rm i}}{2c_s^2 \sqrt{1+\frac{m^2}{\ell_{\rm q}^2}} }\biggr[(-1)^{1+\theta(m)}j_\ell\big(\ell_{\rm q} \mathcal{M}  f^-_m \big)j_{\ell-1}\big(\ell_{\rm q} \mathcal{M}  f^-_m \big)
    \nonumber \\
    & 
    ~~~~~~~~~~~~~~~ +~{\rm i} j_\ell\big( \ell_{\rm q} \mathcal{M}  f^-_m \big)y_{\ell-1}\big(\ell_{\rm q} \mathcal{M}  f^-_m \big) -j_\ell\big( {\rm i} \ell_{\rm q} \mathcal{M}  f^+_m \big)h_{\ell-1}^{(1)}\big( {\rm i} \ell_{\rm q} \mathcal{M}  f^+_m \big)\biggr]\,,
\end{align}
written in terms of the Heaviside step function~$\theta$. By introducing the modified spherical Bessel functions of the first and second kind,~$i_\ell$ and~$k_\ell$, we may remove the imaginary unit from the arguments by exploiting the identities~$j_\ell({\rm i} x)={\rm i}^{\ell} i_\ell(x)$ and~$h^{(1)}_{\ell-1}({\rm i} x)=\frac{2}{\pi}{\rm i}^{-\ell}k_{\ell-1}(x)$. We arrive at
\begin{align}
    S_{\ell,\ell-1}^{m}&= \frac{\pi {\rm i}}{2c_s^2 \sqrt{1+\frac{m^2}{\ell_{\rm q}^2}} }\biggr[(-1)^{1+\theta(m)}j_\ell\big(\ell_{\rm q} \mathcal{M}  f^-_m \big)j_{\ell-1}\big(\ell_{\rm q} \mathcal{M}  f^-_m \big)
    \nonumber \\
    & 
    ~~~~~~~~~~~~~~~ +~{\rm i} j_\ell\big( \ell_{\rm q} \mathcal{M}  f^-_m \big)y_{\ell-1}\big(\ell_{\rm q} \mathcal{M}  f^-_m \big) 
   +\frac{2}{\pi}i_\ell\big( \ell_{\rm q} \mathcal{M}  f^+_m\big)k_{\ell-1}\big( \ell_{\rm q} \mathcal{M}  f^+_m\big)\biggr]\,.
\end{align}
This coincides with Eq.~\eqref{fullFr} in the main text.

\end{appendices}

\bibliographystyle{JHEP}
\bibliography{draft.bib}
\end{document}